\documentclass[preprint,12pt]{article}

\usepackage{amssymb}
\usepackage{amsmath}
\usepackage{amsfonts}
\usepackage{mathtools}
\usepackage{booktabs}

\usepackage{todonotes}

\usepackage{tikz}
\usepackage{tikz-cd}
\tikzcdset{arrow style=tikz, diagrams={>=latex}}
\usepackage{enumerate}

\usepackage{amsthm}

\usepackage{color,colortbl}
\usepackage{tcolorbox}
\usepackage{hyperref}
\usepackage{xspace}
\usepackage{hhline}

\theoremstyle{plain}
\newtheorem{teor}{Theorem}
\newtheorem*{teor*}{Theorem}

\newtheorem{prop}[teor]{Proposition}
\newtheorem{cor}[teor]{Corollary}

\theoremstyle{definition}
\newtheorem{defn}[teor]{Definition}
\newtheorem*{defn*}{Definition}

\theoremstyle{remark}

\newtheorem{rk}[teor]{Remark}

\newtheorem*{claim*}{Claim}
\newtheorem{claim}[teor]{Claim}
\newenvironment{claimproof}
 {\proof} 
 {\endproof}

\newcommand{\RS}{\mathcal{RS}}

\newcommand{\res}{\mathrm{res}}
\newcommand{\en}{\textnormal{en}}

\newcommand{\mN}{\mathbb{N}}

\newcommand{\sA}{\mathcal{A}}
\newcommand{\sB}{\mathcal{B}}
\newcommand{\sfR}{\mathsf{R}}
\newcommand{\sfI}{\mathsf{I}}
\newcommand{\sfP}{\mathsf{P}}
\newcommand{\sfS}{\mathsf{S}}
\newcommand{\sfT}{\mathsf{T}}
\newcommand{\sfU}{\mathsf{U}}
\newcommand{\sfV}{\mathsf{V}}
\newcommand{\SO}{\mathsf{SO}}

\newcommand{\coNP}{\textnormal{\textbf{coNP}}\xspace}
\newcommand{\NP}{\textnormal{\textbf{NP}}\xspace}
\newcommand{\Pp}{\textnormal{\textbf{P}}\xspace}
\newcommand{\pItwoP}{\ensuremath{\mathbf{\Pi}_2^{\Pp}}\xspace}
\newcommand{\SigtwoP}{\ensuremath{\mathbf{\Sigma}_2^{\Pp}}\xspace}

\newcommand{\posV}{\textnormal{pos}}
\newcommand{\negV}{\textnormal{neg}}
\newcommand{\n}[1]{\overline{#1}}

\colorlet{myBlue}{cyan!10!white}
\colorlet{myGray}{lightgray!30!white}

\begin{document}

\author{Rocco Ascone$^\dag$\footnote{Corresponding author} \and Giulia Bernardini$^\dag$ \and Luca Manzoni\footnote{University of Trieste, Italy}}

\date{}

\title{Fixed Points and Attractors of Reactantless and Inhibitorless Reaction Systems}
\maketitle

\begin{abstract}
Reaction systems are discrete dynamical systems that model biochemical processes in living cells using finite sets of reactants, inhibitors, and products.
We investigate the computational complexity of a comprehensive set of problems related to the existence of fixed points and attractors in two constrained classes of reaction systems, in which either reactants or inhibitors are disallowed. These problems have biological relevance and have been extensively studied in the unconstrained case; however, they remain unexplored in the context of reactantless or inhibitorless systems. Interestingly, we demonstrate that although the absence of reactants or inhibitors simplifies the system's dynamics, it does not always lead to a  reduction in the complexity of the considered problems.
\end{abstract}

\section{Introduction}
\label{sec:intro}
Reaction systems are an abstract model of computation inspired by the chemical reactions in living cells, introduced by Ehrenfeucht and Rozenberg almost two decades ago~\cite{DBLP:conf/dlt/EhrenfeuchtR04, DBLP:journals/fuin/EhrenfeuchtR07}. The idea at the basis of reaction systems is that the processes carried out by biochemical reactions within a cell can be described through a finite set of entities, modelling different substances, and a finite set of rules, modelling reactions. A reaction is defined by a set of reactants, a set of inhibitors and a set of products: if any current set of entities (defining a \emph{state} of the reaction system) includes the set of reactants and does not contain any of the inhibitors, the reaction takes place and the products are generated. 

Reaction systems are a \emph{qualitative model}: 
they assume that if a reactant is present at a certain state, then its amount is always enough for all the reactions that use it to take place (i.e., reactions do not conflict even if they share some resources).
The next state of the reaction system is then given by the union of the products of the reactions that took place. 

Despite the simplicity of this formulation, it has been proven that reaction systems can simulate a variety of real-world biological processes~\cite{DBLP:journals/tcs/CorolliMMBM12}, including heat shock response~\cite{DBLP:journals/fuin/AzimiIP14}, gene regulatory networks~\cite{DBLP:journals/fuin/BarbutiBGGLM21} and oncogenic signalling~\cite{DBLP:journals/jmemcom/IvanovP20}.
Studying the computational complexity of the dynamics of reaction systems is also a rich and active research area~\cite{DBLP:journals/nc/FormentiMP15,DBLP:journals/tcs/AzimiG0MPP16,DBLP:journals/tcs/BarbutiGLM16,DBLP:journals/fuin/NobilePSMCMB17,DBLP:journals/iandc/DennunzioFMP19}; while the standard model for reaction systems does not pose any constraints on the number of reactants and inhibitors involved in the reactions, a different line of research is focused on the study of reaction systems with limited resources (i.e., bounding the number of reactants and/or inhibitors involved in any reaction)~\cite{DBLP:journals/ijfcs/EhrenfeuchtMR11,dennunzio2016reachability,DBLP:journals/tcs/Azimi17}.

This work is in the latter vein and studies the computational complexity of deciding on the occurrence of behaviors related to fixed points, attractors and result functions in the special classes of \emph{reactantless} and \emph{inhibitorless} reaction systems. Since reaction systems can be used to model biological processes, these questions have biological relevance: for instance, determining whether fixed points and cycles are present is crucial in modelling gene regulatory networks~\cite{kauffman2004ensemble,bornholdt2008boolean}; and attractors can represent cellular types or states~\cite{DBLP:journals/pieee/ShmulevichDZ02}. 

To the best of our knowledge, the only existing works studying the complexity of dynamical behaviors in reactantless and inhibitorless reaction systems are concerned with the reachability problem~\cite{dennunzio2016reachability} and the evolvability problem~\cite{teh2022evolvability} (the latter work considering a different type of reaction systems, recently introduced by Ehrenfeucht et al.~\cite{ehrenfeucht2017evolving}). This work is thus an important step towards a full understanding of the dynamics in these constrained models.

\begin{table}[t]\resizebox{\textwidth}{!}{
    \centering
    \renewcommand{\arraystretch}{1.1}
    \begin{tabular}{l c|c|c}
     
     \textbf{Problem} 
     &\multicolumn{1}{|c|}{$\RS(\infty,\infty)$}
     &$\RS(0,\infty)$ 
     &$\RS(\infty,0)$ \\
     
     \toprule
    A given state is a fixed point attractor 
    &\multicolumn{1}{|c|}{\NP-c~\cite{formenti2014fixed}}  
    &\cellcolor{myBlue}{\NP-c (Thm.~\ref{teor: reactnatless_T_fixed_att})} 
    &\cellcolor{myBlue}{\NP-c (Thm.~\ref{teor: inibithorless_T_fixed_att})} \\
    
    \hhline{====} 
    $\exists$ fixed point 
    &\multicolumn{1}{|c|}{\NP-c~\cite{formenti2014fixed}}  
    &\cellcolor{myBlue}{\NP-c (Thm.~\ref{teor: reactantless_exist_fix_point})}
    &\Pp~\cite{Granas2003} \\
    
    \hline
    $\exists$ common fixed point
    &\multicolumn{1}{|c|}{\NP-c~\cite{formenti2014fixed}}  
    &\cellcolor{myBlue}{\NP-c (Cor.~\ref{cor: reactantless_common_fix_point})}
    &\cellcolor{myBlue}{\NP-c (Thm.~\ref{teor: inhibitorless_common_fix_point})} \\
    
    \hline
    sharing all fixed points 
    &\multicolumn{1}{|c|}{\coNP-c~\cite{formenti2014fixed}}  
    &\cellcolor{myBlue}{\coNP-c (Thm.~\ref{teor: reactantless_all_fix_point})}
    &\cellcolor{myBlue}{\coNP-c (Thm.~\ref{teor: inhibitorless_all_fix_point})} \\

    \hhline{====}
    $\exists$ fixed point attractor 
    &\multicolumn{1}{|c|}{\NP-c~\cite{formenti2014fixed}}  
    &\cellcolor{myBlue}{\NP-c (Thm.~\ref{teor: reactantless_fix_point_attr})} 
    &{Unknown} \\
    
    \hline
    $\exists$ common fixed point attractor 
    &\multicolumn{1}{|c|}{\NP-c~\cite{formenti2014fixed}}  
    &\cellcolor{myBlue}{\NP-c (Cor.~\ref{cor: reactantless_common_fix_point_attr})} 
    &\cellcolor{myBlue}{\NP-c (Cor.~\ref{cor: inhibitorless_common_fix_point_attr})} \\
    
    \hline
    sharing all fixed points attractors 
    &\multicolumn{1}{|c|}{\pItwoP-c~\cite{formenti2014fixed}} 
    &\cellcolor{myBlue}{\pItwoP-c (Cor.~\ref{cor: reactantless_all_fix_point_attr})} 
    &\cellcolor{myBlue}{\pItwoP-c (Thm.~\ref{teor: inhibitorless_all_fix_point_attr})} \\ 
    
    \hhline{====} 
    $\exists$ fixed point not attractor  
    & \multicolumn{1}{|c|}{\cellcolor{myBlue}{\SigtwoP-c} (Cor.~\ref{cor: fix_point_not_attr})}
    &\cellcolor{myBlue}{\SigtwoP-c (Thm.~\ref{teor: reactantless_fix_point_not_attr})}
    &\cellcolor{myBlue}{\SigtwoP-c (Cor.~\ref{cor: inhibitorless_fix_point_not_attr})} \\

    \hline
    $\exists$ common fixed point not attractor  
    & \multicolumn{1}{|c|}{\cellcolor{myBlue}{\SigtwoP-c (Cor.~\ref{cor: common_fix_not_attr})}}
    &\cellcolor{myBlue}{\SigtwoP-c (Cor.~\ref{cor: reactantless_common_fix_not_attr})}
    &\cellcolor{myBlue}{\SigtwoP-c (Cor.~\ref{cor: inhibitorless_common_fix_point_not_attr})} \\
    
    \hline
    sharing all fixed points not attractors  
    &\multicolumn{1}{|c|}{\cellcolor{myBlue}{\coNP-c (Cor.~\ref{cor: all_fix_pointge})}}
    & \cellcolor{myBlue}{\coNP-c (Cor.~\ref{cor: reactantless_all_fix_pointge})}
    &\cellcolor{myBlue}{\coNP-c (Cor.~\ref{cor: inhibitorless_all_fix_pointge})} \\
    
    \hhline{====}
    $\res_{\sA} = \res_{\sB}$ 
    &\multicolumn{1}{|c|}{\cellcolor{myBlue}{\coNP-c (Thm.~\ref{teor: res_A=res_B})}}
    &\cellcolor{myBlue}{\Pp(Cor.~\ref{cor: reactantless_resA=resB})} 
    &\cellcolor{myBlue}{\Pp (Cor.~\ref{cor: inibithorless_resA=resB})} \\
    
    \hline
    res bijective 
    &\multicolumn{1}{|c|}{\coNP-c~\cite{formenti2014cycles}}  
    &\cellcolor{myBlue}{\Pp (Cor.~\ref{cor: res_A_bijective_reactantless})} 
    &\cellcolor{myBlue}{\Pp (Cor.~\ref{cor: res_A_bijective_inhibitorless})} \\ 
    
    \hline
    \end{tabular}}
    \caption{Computational complexity of the problems studied in this work for different classes of reaction systems. \NP-c, \coNP-c, \SigtwoP-c and \pItwoP-c are shorthands for \NP-complete, \coNP-complete, \SigtwoP-complete and \pItwoP-complete, respectively; $\RS(\infty,\infty)$, $\RS(0,\infty)$ and $\RS(\infty,0)$ denote unconstrained, reactantless and inhibitorless reaction systems, respectively (see Def.~\ref{def:classes}). Light-blue cells contain the results proved in this paper.}
    \label{tab: problems_complexity}
\end{table}

The main contributions of this study are summarized in Table~\ref{tab: problems_complexity} (light-blue cells). It is interesting to notice that, although the dynamical behaviors of the resource-bounded systems are less rich than those of unrestricted systems, the complexities of the considered problems are not necessarily reduced. For example, the complexity of deciding whether two reaction systems have a common fixed point is \NP-complete in the general case and it remains so in both the resource-bounded classes we consider. 
In contrast, e.g., deciding whether the result function is bijective is NP-complete in the unconstrained case, while it can be done in polynomial time in reactantless and inhibitorless systems.

Furthermore, the complexity of a problem in reactantless reaction systems is not always the same as in inhibitorless systems: e.g., deciding on the existence of a fixed point can be done in polynomial time in inhibitorless reaction systems, while it is an \NP-complete problem in reactantless systems. The complexities of different problems for different classes of reaction systems thus vary in a non-uniform way.

This paper is organized as follows. In Section~\ref{sec:basics}, we provide basic notions on reaction systems 
and introduce the notation we use throughout the paper. In Section~\ref{sec:logic} we report a description of reaction systems and their dynamics in terms of logical formulae which will be useful to prove most of our results. 
In Section~\ref{sec: T_fix_attr}, we study the problem of deciding whether a given state is a fixed point attractor in both constrained models.
In Section~\ref{sec: fix_point_0_infty} we study the complexities of fixed point problems in reactantless systems; in Section~\ref{sec: fix_point_infty_0} we study the same problems in inhibitorless systems. In Section~\ref{sec:resultFunction} we study the problem of deciding whether two reaction systems have the same result function both in the general model and in the constrained classes. In Section~\ref{sec: bij_res_A} we consider the problem of deciding whether the result function is bijective both in reactantless and inhibitorless systems.  Finally, in Section~\ref{sec:conclusions} we discuss our results and suggest future research directions.

\section{Basics Notions}\label{sec:basics}

In this section, we introduce the notation used in the paper and the main definitions concerning reaction systems.

Given a finite set $S$ of \emph{entities}, a \emph{reaction} $a$ over $S$ is a triple $( R_a,I_a,P_a)$ of subsets of $S$\footnote{While we do not restrict $R_a$ and $I_a$ to be disjoint, as it is usually done, it should be clear that all the results of the paper hold even then this assumption holds.}; the set $R_a$ is the set of \emph{reactants}, $I_a$ is the set of \emph{inhibitors}, and $P_a$ is the nonempty set of \emph{products}. 
We remark that in this paper the set of reactants and inhibitors of a reaction are allowed to be empty as in the original definition \cite{DBLP:conf/dlt/EhrenfeuchtR04}.
The set of all reactions over $S$ is denoted by $\text{rac}(S)$.
A \emph{reaction system} (RS) is a pair $\sA = (S,A)$ where $S$ is a finite set of entities, called the \emph{background set}, and $A \subseteq \text{rac}(S)$.

Given a \emph{state} $T \subseteq S$, a reaction $a$ is said to be \emph{enabled} in T when $R_a \subseteq  T$ and $I_a \cap T = \varnothing$.
The \emph{result function} $\res_a: 2^S \to 2^S$ of $a$, where $2^S$
denotes the power set of $S$, is defined as 
\begin{equation*}
    \res_a(T) \coloneqq
    \begin{cases}
        P_a & \text{if $a$ is enabled by $T$} \\
        \varnothing & \text{otherwise}.
    \end{cases}
\end{equation*}
The definition of $\res_a$ naturally extends to sets of reactions.
Indeed, given $T \subseteq S$ and $A\subseteq \text{rac}(S)$, define $\res_{A} (T)\coloneqq \bigcup_{a \in A} \res_a(T)$.
The result function $\res_{\sA}$ of a RS $\sA = (S, A)$ is defined to be equal to $\res_A$, i.e., the result function on the whole set of reactions. 

In this way, any RS $\sA = (S, A)$ induces a discrete dynamical system where the state set is $2^S$ and the next state function is $\res_{\sA}$. 

In this paper, we are interested in the dynamics of RS, i.e., the study of the successive states of the system under the action of the result function $\res_{\sA}$ starting from some initial set of entities.
The set of reactions of $\sA$ enabled in a state $T$ is denoted by $\en_{\sA}(T)$.
The \emph{orbit} or \emph{state sequence} of a given state $T$ of a RS $\sA$ is defined as the sequence of states obtained by subsequent iterations of $\res_{\sA}$ starting from $T$, namely the sequence $(T,\res_{\sA}(T),\res_{\sA}^2(T), \dots)$.
We remark that since $S$ is finite, for any state $T$ the sequence $(\res_{\sA}^n(T))_{n \in \mN}$ is ultimately periodic.

Given a RS $\sA$ with background set $S$, a \emph{fixed point} $T\subseteq S$ is a state such that $\res_{\sA}(T) = T$.
A \emph{fixed point attractor} is a fixed point $T$ for which there exists a state $U\ne T$ such that $\res_{\sA}(U) = T$.
A \emph{fixed point not attractor} is a fixed point that is not an attractor, i.e., not reachable from any state other than $T$ itself.

We now recall the classification of reaction systems in terms of the number of resources employed per reaction \cite{manzoni2014simple}.

\begin{defn}\label{def:classes}
    Let $i,r\in \mN$. The class $\RS(r,i)$ consists of all RS having at most $r$ reactants and $i$ inhibitors for reaction. 
    We also define the (partially) unbounded classes $\RS(\infty,i) = \bigcup_{r = 0}^\infty \RS(r,i)$, $\RS(r,\infty) = \bigcup_{i = 0}^\infty \RS(r,i)$, and $\RS(\infty, \infty) = \bigcup_{r= 0}^{\infty}\bigcup_{i = 0}^\infty \RS(r,i)$.
\end{defn}

In the following, we will call $\RS(0,\infty)$ the class of \emph{reactantless} systems, and $\RS(\infty,0)$ the class of \emph{inhibitorless} systems.

\begin{defn}[\cite{manzoni2014simple}]
    Let $\sA = (S, A)$ and $\sA' = (S', A)$, with $S \subseteq S'$, be two reaction systems, and let $k \in \mN$. 
    We say that $\sA'$ \emph{$k$-simulates} $\sA$ if and only if, for all $T \subseteq S$ and all $n \in \mN$, we have
    \begin{equation*}
        \res_{\sA}^n(T) = \res_{\sA'}(T)^{kn} \cap S.
    \end{equation*}
\end{defn}

\begin{defn}[\cite{manzoni2014simple}]
    Let $X$ and $Y$ be classes of reaction systems, and let $k \in \mN$.
    We define the binary relation $\preceq_k$ as follows: $X \preceq_k Y$ if and only if for all $\sA \in X$ there exists a reaction system in Y that $l$-simulates $\sA$ for some $l\le k$.
    We say that $X \preceq Y$ if and only if $X \preceq_k Y$ for some $k \in \mN$. 
    We write $X \approx_k Y$ if $X \preceq_k Y$ and $Y \preceq_k X$, and $X \approx Y$ for $X \preceq Y \land Y \preceq X$.
    Finally, the notation $X \prec Y$ is shorthand for $X \preceq Y \land Y \npreceq X$.
\end{defn}

The relation $\preceq$ is a preorder and the relation $\approx$ induces exactly five equivalence classes \cite[Theorem 30]{manzoni2014simple}:
\begin{equation}
    \label{eq: equiv_classes}
    \RS(0,0) \prec \RS(1,0) \prec \RS(\infty,0) \prec \RS(0,\infty) \prec \RS(\infty, \infty).
\end{equation}

We remark that this classification does not include the number of products as a parameter because RS can always be assumed to be in \emph{singleton product normal form} \cite{brijder2011reaction}: any reaction $(R, I, \{p_1,\dots, p_m\})$ can be replaced by the set of reactions $(R, I, \{p_1\}),\dots,(R, I, \{p_m\})$ since they produce the same result.

The five equivalence classes in (\ref{eq: equiv_classes}) have a characterisation in terms of functions over the Boolean lattice $2^S$ \cite{manzoni2014simple}, see Table~\ref{tab: class&function}.
\begin{table}
    \centering
    \begin{tabular}{ll}
         \textbf{Class of RS} & \textbf{Subclass of $2^S\to 2^S$} \\
         \hline
         $\RS(\infty,\infty)$   & all \\
         $\RS(0,\infty)$        & antitone \\
         $\RS(\infty,0)$        & monotone \\
         $\RS(1,0)$             & additive \\
         $\RS(0,0)$             & constant
    \end{tabular}
    \caption{Functions computed by restricted classed of RS.}
    \label{tab: class&function}
\end{table}
Recall that a function $f : 2^S \to 2^S$ is \emph{antitone} if $X \subseteq Y$ implies $f(X) \supseteq f(Y)$, \emph{monotone} if $X \subseteq Y$ implies $f(X) \subseteq f(Y)$, \emph{additive} (or an \emph{}{upper-semilattice endomorphism}) if $f(X \cup Y ) = f(X) \cup f(Y)$ for all $X,Y \in 2^S$.
We say that the RS $\sA = (S, A)$ computes the function $f : 2^S \to 2^S$ if $\res_{\sA} = f$.

\section{Logical Description}\label{sec:logic}

In this section, we recall a logical description of RS and formulae related to their dynamics (see \cite{formenti2014fixed} for its first introduction).
This description is sufficient for proving membership in many complexity classes. For the background notions of logic and descriptive complexity, we refer the reader to the classical book of Neil Immerman \cite{immerman1999descriptive}.

Each of the problems studied in this work can be characterised by a logical formula.
A RS $\sA = (S, A)$ with background set $S \subseteq \{0,\dots, n-1\}$ and $|A| \le n$ can be described by the vocabulary $(\sfS, \sfR_{\sA}, \sfI_{\sA}, \sfP_{\sA})$, where $\sfS$ is a unary relation symbol and $\sfR_{\sA}, \sfI_{\sA}, \sfP_{\sA}$ are binary relation symbols.
The intended meaning of the symbols is the following: the set of entities is $S = \{i : \sfS(i)\}$ and each reaction $a_j = (R_j , I_j , P_j ) \in A$ is described by the sets $R_j = \{i \in S : \sfR_{\sA}(i, j)\}$, $I_j = \{i \in S : \sfI_{\sA}(i, j)\}$, and $P_j = \{i \in S : \sfP_{\sA}(i, j)\}$.
We will also need some additional vocabularies: $(\sfS, \sfR_{\sA}, \sfI_{\sA}, \sfP_{\sA},\sfT)$, where $\sfT$ is a unary relation representing a subset of $S$, $(\sfS, \sfR_{\sA}, \sfI_{\sA}, \sfP_{\sA},\sfT_1,\sfT_2)$ with two additional unary relations representing sets, and $(\sfS, \sfR_{\sA}, \sfI_{\sA}, \sfP_{\sA}, \sfR_{\sB}, \sfI_{\sB}, \sfP_{\sB})$ denoting two RS's over the same background set.

The following formulae, introduced in~\cite{formenti2014fixed}, describe the basic properties of $\sA$. 
The first is true if a reaction $a_j$ is enabled in $T$:
\begin{equation*}
    \textsc{en}_{\sA}(j, \sfT) \equiv \forall i(\sfS(i) \Rightarrow (\sfR_{\sA}(i, j) \Rightarrow \sfT(j)) \land (\sfI_{\sA}(i, j) \Rightarrow \neg \sfT(j)))
\end{equation*}
and the following is verified if $\res_{\sA}(T_1) = T_2$ for $T_1, T_2 \subseteq S$:
\begin{equation*}
    \textsc{res}_{\sA}(\sfT_1,\sfT_2) \equiv \forall i(\sfS(i) \Rightarrow (\sfT_2(i) \Leftrightarrow \exists j(\textsc{en}_{\sA}(j, \sfT_1) \land \sfP_{\sA}(i, j))).
\end{equation*}
Notice that $\textsc{en}_{\sA}$ and $\textsc{res}_{\sA}$ are both first-order ($\mathsf{FO}$) formulae.
We define the bounded second-order quantifiers $(\forall X \subseteq Y) \varphi$ and $(\exists X \subseteq Y) \varphi$ as a short-hand for $\forall X (\forall i(X(i) \Rightarrow Y(i)) \Rightarrow \varphi)$ and $\exists X (\forall i(X(i) \Rightarrow Y(i)) \land  \varphi)$.
We will use the following formulae (given in~\cite{formenti2014fixed}) to describe our problems:
\begin{align*}
    \textsc{fix}_{\sA}(\sfT) &\equiv \textsc{res}_{\sA}(\sfT,\sfT)
    \\
    \textsc{reach}_{\sA}(\sfT) &\equiv (\exists \sfU \subseteq \sfS) (\textsc{res}_{\sA}(\sfU,\sfT) \land \neg\textsc{res}_{\sA}(\sfT,\sfU)) 
    \\
    \textsc{att}_{\sA}(\sfT) &\equiv \textsc{fix}_{\sA}(\sfT) \land \textsc{reach}_{\sA}(\sfT)  
    \\
    \textsc{fixge}_{\sA}(\sfT) &\equiv \textsc{fix}_{\sA}(\sfT) \land \neg\textsc{reach}_{\sA}(\sfT)  
    \\    
    \textsc{res\textunderscore eq}_{\sA,\sB} &\equiv (\forall \sfT \subseteq \sfS) (\forall \sfV\subseteq \sfS)(\textsc{res}_{\sA}(\sfT,\sfV) \Leftrightarrow \textsc{res}_{\sB}(\sfT,\sfV)).
\end{align*}

We say that a formula is $\SO\exists$, $\SO\forall$, or $\SO\forall\exists$ if it is logically equivalent to a formula in the required prenex normal form.

In Table~\ref{tab: problem_logic_class}, we give the logic formulae associated with the problems considered in this work. The first six formulae were already given in~\cite{formenti2014fixed}; we added the last four to describe the new problems studied in this paper.
\begin{table}[t]\resizebox{\textwidth}{!}{
    \centering
    \renewcommand{\arraystretch}{1.1}
    
    \begin{tabular}{|l|l|c|}
    \hline
    \textbf{Problem} & \textbf{Formula} & \textbf{Logic class} \\
     
     \hline
    $\exists$ fixed point & $(\exists \sfT \subseteq \sfS) \textsc{fix}_{\sA}(\sfT)$ &$\SO\exists$ \\

    \hline
    $\exists$ common fixed point &$(\exists \sfT \subseteq \sfS) (\textsc{fix}_{\sA}(\sfT) \land \textsc{fix}_{\sB}(\sfT))$ &$\SO\exists$ \\
    
    \hline
    sharing all fixed points  & $(\forall \sfT \subseteq \sfS) (\textsc{fix}_{\sA}(\sfT) \Leftrightarrow \textsc{fix}_{\sB}(\sfT))$ & $\SO\forall$ \\
    
    \hline
    A given state is a fixed point attractor  & $\textsc{att}_{\sA}(\sfT)$ & $\SO\exists$ \\
    
    \hline
    $\exists$ fixed point attractor  & $(\exists \sfT \subseteq \sfS) \textsc{att}_{\sA}(\sfT)$ &  $\SO\exists$ \\
    
    \hline
    $\exists$ common fixed point attractor  & $(\exists \sfT \subseteq \sfS) \textsc{att}_{\sA}(\sfT) \land \textsc{att}_{\sB}(\sfT)$ & $\SO\exists$ \\
    
    \hline
    sharing all fixed points attractors  & $(\forall \sfT \subseteq \sfS) (\textsc{att}_{\sA}(\sfT) \Leftrightarrow \textsc{att}_{\sB}(\sfT))$ & $\SO\forall\exists$ \\
    
    \hline
    $\exists$ fixed point not attractor  
    &  $(\exists \sfT \subseteq \sfS) \textsc{fixge}(\sfT)$ 
    & $\SO\exists\forall$ \\

    \hline
    $\exists$ common fixed point not attractor  
    & $(\exists \sfT \subseteq \sfS) (\textsc{fixge}_{\sA}(\sfT) \land \textsc{fixge}_{\sB}(\sfT))$ 
    &$\SO\exists\forall$ \\
    
    \hline
    sharing all fixed points not attractors  &
    $(\forall \sfT \subseteq \sfS) (\textsc{fixge}_{\sA}(\sfT) \Leftrightarrow \textsc{fixge}_{\sB}(\sfT))$ &
    $\SO\forall$ \\
    
    \hline
    $\res_{\sA} = \res_{\sB}$  & $\textsc{res\textunderscore eq}_{\sA,\sB}$ & $\SO\forall$ \\
    
    \hline
    \end{tabular}}
    \caption{Problems with the associated formula and logic class.}
    \label{tab: problem_logic_class}
\end{table}

Remark that existential second-order logic $\SO\exists$ characterizes \NP (Fagin’s theorem~\cite{immerman1999descriptive});
universally quantified second-order logic $\SO\forall$ gives \coNP; second-order logic with one alternation of existential and universal quantifiers $\SO\exists\forall$ gives \SigtwoP, and, in a dual way, second-order logic with one alternation of universal and existential quantifiers $\SO\forall\exists$ gives \pItwoP~\cite{immerman1999descriptive}.

The hardness proofs we give in this paper are obtained via reductions from \textsc{sat}, \textsc{validity}~\cite{papadimitriou1994computational} or $\forall\exists$\textsc{sat}~\cite{STOCKMEYER19761}. In Definition~\ref{def:sat_entities} we introduce a few sets of entities corresponding to a Boolean formula in CNF or DNF that will be used in such proofs.

\begin{defn}\label{def:sat_entities}
    Let $\varphi$ be a Boolean formula in CNF or DNF with clauses $C=\{\varphi_1,\dots,\varphi_m\}$ over the variables $V = \{x_1,\dots,x_n\}$. In the reductions, we will consider $V$ as a set of entities representing true assignments to the variables and $C$ as a set of entities representing the clauses. We also define the following additional sets of entities. 
    
    $\n{V}\coloneqq\{\n{x_j}: x_j \in V\}$, entities representing false assignments to the variables; $\posV(\varphi_r) \subseteq V$, representing the set of variables that occur non-negated in $\varphi_r$ to which is assigned a true value; $\negV(\varphi_r) \subseteq V$, representing the set of variables that occur negated in $\varphi_r$ to which is assigned a true value; $\n\posV(\varphi_r) \subseteq \n{V}$, representing the set of variables that occur non-negated in $\varphi_r$ to which is assigned a false value; and $\n\negV(\varphi_r) \subseteq \n{V}$, representing the set of variables that occur negated in $\varphi_r$ to which is assigned a false value.

    In addition, we will use card suits symbols to denote extra entities that do not belong to any of the sets defined above.
\end{defn}

In general, in the reductions we will adopt the convention that overlining a set of variables means that those variables are assigned a false value. Indeed, the elements of the power set of $V$ and $\n{V}$ are in a natural bijection: thus 
for any $X\subseteq V$, $\n{X}$ denotes the assignment of a false value to the variables in $X$.
In particular, we can represent any assignment for $\varphi$ as a subset of entities $X \cup \n{V \setminus X}$, where $X\subseteq V$ is the subset of variables that are set to true. Furthermore, $X \cup \n{V \setminus X} \vDash \varphi$ (resp.$X \cup \n{V \setminus X} \nvDash \varphi$) means that the assignment satisfies (resp. does not satisfy) $\varphi$.

\section{A Given State is a Fixed Point Attractor}
\label{sec: T_fix_attr}
In this section, we study the problem of deciding whether a given state is a fixed point attractor.
The problem is \NP-complete for $\RS(\infty,\infty)$ \cite[Theorem 4]{formenti2014fixed}; we prove that it remains so also for reactantless and inhibitorless reaction systems.

\begin{teor}
    \label{teor: inibithorless_T_fixed_att}
    Given $\sA\in \RS(\infty,0)$ and a state $T \subseteq S$, it is \NP-complete to
    decide whether $T$ is a fixed point attractor.
    \begin{proof}
	The problem is in \NP (see Table~\ref{tab: problem_logic_class}). 
	In order to show \NP-hardness, we give a reduction from \textsc{sat} \cite{papadimitriou1994computational}. 
        Given a Boolean formula $\varphi = \varphi_1 \land \cdots \land \varphi_m$ in CNF over the variables $V = \{x_1,\dots,x_n\}$, we define a RS $\sA$ with background set $S\coloneqq V \cup \n{V} \cup C \cup \{\spadesuit\}$, where $C,V,\n{V}, \spadesuit$ are as in Definition~\ref{def:sat_entities}, and reactions 
        \begin{align}
            \label{eq: x,empty,varphi}
            &(\{x\},\varnothing,\{\varphi_j\})                   & \text{ for } &1\le j \le m \text{ and } x \in \posV(\varphi_j)\\
            \label{eq: negx,empty,varphi}
            &(\{\n{x} \},\varnothing,\{\varphi_j\})	             & \text{ for } &1\le j \le m \text{ and } \n{x} \in \n\negV(\varphi_j)\\
            \label{eq: x_i,neg_x_i,empty,spade}
            &(\{x_i,\n{x_i}\},\varnothing,\{\spadesuit\}) 	     & \text{ for } &1\le i \le n\\
            \label{eq: x_i,varphi,empty,spade}
            &(\{x_i,\varphi_j\},\varnothing,\{\spadesuit\})      & \text{ for } &1\le i \le n \text{ and } 1\le j \le m\\
            \label{eq: neg_x_i,varphi,empty,spade}
            &(\{\n{x_i},\varphi_j\},\varnothing,\{\spadesuit\})  & \text{ for } &1\le i \le n \text{ and } 1\le j \le m\\
            \label{eq: varphi,empty,varphi}
            &(\{\varphi_j\},\varnothing,\{\varphi_j\})           & \text{ for } &1\le j \le m \\
            \label{eq: spade,empty,spade}
            &(\{\spadesuit\},\varnothing,\{\spadesuit\}).
        \end{align}
	Reactions (\ref{eq: varphi,empty,varphi}) imply that $T\coloneqq C$ is a fixed point for $\sA$.
	Furthermore, consider a state $T' \ne C$ such that $\res_{\sA}(T') \subseteq C$. It must hold $T' = X_1 \cup \n{X_2}$ with $X_1 \subseteq V$, $X_2\subseteq V$ and $X_1 \cap X_2 = \varnothing$, as otherwise $\spadesuit$ would be generated by one of the reactions of type (\ref{eq: x_i,neg_x_i,empty,spade}), (\ref{eq: x_i,varphi,empty,spade}), (\ref{eq: neg_x_i,varphi,empty,spade}) or (\ref{eq: spade,empty,spade}).
	Furthermore, consider $\varphi_j\in \res_{\sA}(T')$: since $T' \cap C = \varnothing$, then either $X_1 \cap \posV(\varphi_j) \ne \varnothing$ or $X_2 \cap \negV(\varphi_j) \ne \varnothing$ because at least one of the reactions of type (\ref{eq: x,empty,varphi}) or (\ref{eq: negx,empty,varphi}) with $\varphi_j$ in the product must be enabled.
	We can thus interpret $T'$ as an assignment satisfying $\varphi$.
		
        We remark that if $X_1 \cup X_2 \subsetneq V$, there must be some variables that are irrelevant to the satisfiability of $\varphi$.
	Therefore, we proved that if $C$ is an attractor then $\varphi$ is satisfiable.
	The converse follows immediately, since if $X \subseteq V$ is the subset of variables that are given a true value in an assignment that satisfies $\varphi$, then $X \cup \n{V\setminus X}$ enables at least one reaction of type  (\ref{eq: x,empty,varphi}) or (\ref{eq: negx,empty,varphi}) for each clause, thus $\res_{\sA} (X \cup \n{V\setminus X}) = C$.
	We obtain that $T=C$ is a fixed point attractor if and only if $\varphi$ is satisfiable.
	The mapping $\varphi \mapsto (\sA, T)$ is clearly computable in polynomial time, hence deciding if a given fixed point $T$ is an attractor is \NP-hard.
	\end{proof}
\end{teor}

We highlight that all the reactions used in the proof of Theorem~\ref{teor: inibithorless_T_fixed_att} involve at most two reactants, implying that the problem is \NP-complete even for the more restricted class $\RS(2,0)$. 
Note, however, that if we are given a fixed point of $\sA$, it is in \Pp to decide whether it is a special kind of attractor: namely if it can be reached from one of its subsets or from one of its supersets, as we prove in the following remark.

\begin{rk} 
    \label{rk: T-x_Tux}
    Let $\sA=(S,A) \in \RS(\infty,0)$ and $T$ be a fixed point for $\res_{\sA}$. If there exists $T'\subsetneq T$ such that $\res_{\sA}(T') = T$, then for all $x \in T\setminus T'$ 
        \begin{equation*}
            T = \res_{\sA}(T') \subseteq \res_{\sA}(T \setminus \{x\}) \subseteq \res_{\sA}(T) = T,
        \end{equation*}
        thus $T$ is reachable from $T\setminus \{x\}$; and if there exists $T''\supsetneq T$ such that $\res_{\sA}(T'') = T$, then for all $x \in T''\setminus T$ 
        \begin{equation*}
            T = \res_{\sA}(T) \subseteq \res_{\sA}(T \cup \{x\}) \subseteq \res_{\sA}(T'') = T,
        \end{equation*}
        thus $T$ is reachable from $T\cup \{x\}$. To decide whether $T$ is an attractor of the form $T'$ and $T''$ it thus suffices to check, for all $x\in S$, whether $T$ is reachable from $T\setminus \{x\}$ or $T\cup\{x\}$, and it thus requires polynomial time.
\end{rk}

\begin{figure}
    \centering
     \includegraphics{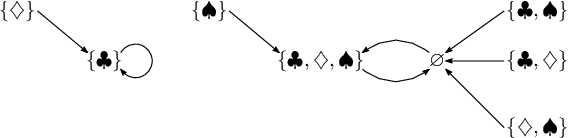}
     \caption{Graph representation of the dynamics given by the reactions (\ref{eq: empty,diamondspade,club}), (\ref{eq: empty,clubspade,club}), (\ref{eq: empty,clubdiamond,clubdiamondspade}) over the subset of the background set $\{\clubsuit,\diamondsuit,\spadesuit\}$.}
     \label{fig: dynamics_club-diamond-spade}
\end{figure}

\begin{teor}
    \label{teor: reactnatless_T_fixed_att}
    Given $\sA\in \RS(0,\infty)$ and a state $T \subseteq S$, it is \NP-complete to
    decide whether $T$ is a fixed point attractor.
    \begin{proof}
	The problem is in \NP (see Table~\ref{tab: problem_logic_class}). 
	In order to show \NP-hardness, we give a reduction from \textsc{sat} \cite{papadimitriou1994computational}. Given a Boolean formula $\varphi = \varphi_1 \land \cdots \land \varphi_m$ in CNF over the variables $V = \{x_1,\dots,x_n\}$, we define $\sA$ a RS with background set $S\coloneqq V \cup \n{V} \cup C \cup \{\clubsuit,\diamondsuit,\spadesuit\}$, where each set is as defined in Definition~\ref{def:sat_entities}, and the set of reactions is
	\begin{align}
            \label{eq: empty,negx,varphi}
            &(\varnothing,\{\n{x},\clubsuit,\spadesuit\},\{\varphi_j\})  			& \text{ for } &1\le j \le m \text{ and } \n{x} \in \n\posV(\varphi_j) \\
            \label{eq: empty,x,varphi}
            &(\varnothing,\{x,\clubsuit,\spadesuit\},\{\varphi_j\})	& \text{ for } &1\le j \le m \text{ and } x \in \negV(\varphi_j)\\
            \label{eq: empty,x negx,spade}
            &(\varnothing,\{ x_i,\n{x_i},\clubsuit,\spadesuit\},\{\spadesuit\}) 		& \text{ for } &1\le i \le n\\
            \label{eq: empty,varphi,spade}
            &(\varnothing,\{\varphi_j\},\{\spadesuit\})  				& \text{ for } &1\le j \le m \\
            \label{eq: C_club_fixed}
            &(\varnothing, S\setminus(C\cup \{\clubsuit\}),C) 		\\
            \label{eq: empty,diamondspade,club}
            &(\varnothing,\{\diamondsuit,\spadesuit\},\{\clubsuit\})\\
            \label{eq: empty,clubspade,club}
            &(\varnothing,\{\clubsuit,\spadesuit\},\{\clubsuit\})\\
            \label{eq: empty,clubdiamond,clubdiamondspade}
            &(\varnothing,\{\clubsuit,\diamondsuit\},\{\clubsuit,\diamondsuit,\spadesuit\})
        \end{align}
        By reactions (\ref{eq: C_club_fixed}) and (\ref{eq: empty,diamondspade,club}), $T \coloneqq C\cup \{\clubsuit\}$ is a fixed point for $\sA$.
	Let $T'\ne C \cup \{\clubsuit\}$ such that $\res_{\sA}(T') = C \cup \{\clubsuit\}$, then $C \subseteq T'$, otherwise $\spadesuit \in \res_{\sA}(T')$ because of the reactions of type (\ref{eq: empty,varphi,spade}).
	In order to have $\res_{\sA}(T') \cap \{\clubsuit,\diamondsuit,\spadesuit\} = \{\clubsuit\}$, $T'$ must either satisfy $T' \cap \{\clubsuit,\diamondsuit,\spadesuit\} = \{\clubsuit\}$ or  $T' \cap \{\clubsuit,\diamondsuit,\spadesuit\} = \{\diamondsuit\}$ (see Figure \ref{fig: dynamics_club-diamond-spade}). We prove by contradiction that the first case cannot occur: indeed, in this case, the only reactions enabled by $T'$ are (\ref{eq: C_club_fixed}) and (\ref{eq: empty,diamondspade,club}). 
	Since adding any element of $V \cup \n{V}$ to $T'$ would disable (\ref{eq: C_club_fixed}) and we have $C \subseteq \res_{\sA}(T')$, we deduce that we would have $T' = C \cup \{\clubsuit\}$, a contradiction.
	Therefore $T'$ must be of the form $X_1 \cup \n{X_2} \cup C \cup  \{\diamondsuit\}$, where $X_1\subseteq V$ and $\n{X_2}\subseteq \n{V}$.
	If we had $X_1 \cup X_2 \subsetneq V$ then we would also have $\spadesuit\in \res_{\sA}(T')$ because one of the reactions (\ref{eq: empty,x negx,spade}) would be enabled; thus it must hold $X_1 \cup X_2 = V$.
  
	If $x \in X_1 \cap X_2$ then neither $x$ nor $\n{x}$ will be able to generate any $\varphi_j$ because of reactions (\ref{eq: empty,negx,varphi}) and (\ref{eq: empty,x,varphi}); however, since $C \subseteq \res_{\sA}(T')$, for each $\varphi_j\in C$ there must exist a $x \in V$ such that either $x\in X_1$ or $\n{x}\in \n{X_2}$ satisfies $\varphi_j$.
	Therefore, we proved that if $C \cup \{\clubsuit\}$ is an attractor then $\varphi$ is satisfiable.
	The converse follows immediately, since if $X \subseteq V$ are the variables set to true in an assignment satisfying $\varphi$, then $X \cup \n{V\setminus X} \cup C \cup \{\diamondsuit\}$ is a state attracted by $C \cup \{\clubsuit\}$.
	We obtain that $T=C\cup \{\clubsuit\}$ is a fixed point attractor if and only if $\varphi$ is satisfiable.
	The mapping $\varphi \mapsto (\sA, T)$ is computable in polynomial time, hence deciding if a given fixed point $T$ is an attractor is \NP-hard.
    \end{proof}
\end{teor}

\section{Fixed Points for Reactantless RS}
\label{sec: fix_point_0_infty}

In this section, we prove \NP-hardness, \coNP-hardness, \SigtwoP-hardness and \pItwoP-hardness for problems of fixed points in the class of reactantless RS. 

The problem of deciding if there exists a fixed point is \NP-complete for $\RS(\infty,\infty)$ \cite[Theorem 2]{formenti2014fixed}; the following theorem shows that it remains difficult also in $\RS(0,\infty)$.

\begin{teor}
    \label{teor: reactantless_exist_fix_point}
    Given $\sA \in \RS(0,\infty)$, it is \NP-complete to decide if $\sA$ has a fixed point.
    \begin{proof}
        The problem is in \NP (see Table~\ref{tab: problem_logic_class}). 
        In order to show \NP-hardness, we reduce \textsc{sat} \cite{papadimitriou1994computational} to this problem. Given a Boolean formula $\varphi = \varphi_1 \land \cdots \land \varphi_m$ in CNF over the variables $V = \{x_1,\dots,x_n\}$, we define a RS $\sA$ with background set $S\coloneqq V \cup \n{V} \cup \{\spadesuit\} \cup \{\clubsuit\}$ (the sets are as in Definition~\ref{def:sat_entities}) and the reactions 
        \begin{align}
            \label{eq: n-neg_pos_spade}
            &(\varnothing, \n\negV(\varphi_j) \cup \posV(\varphi_j), \{\spadesuit\}) & \text{ for } &1\le j \le m\\
            \label{eq: 0,x_i,neg_x_i}
            \n{a_i}\coloneqq&(\varnothing,\{x_i\}, \{\n{x_i}\}) & \text{ for } &1\le i \le n\\
            \label{eq: 0,neg_x_i,x_i}
            a_i\coloneqq&(\varnothing,\{\n{x_i}\}, \{x_i\}) & \text{ for } &1\le i \le n\\
            \label{eq: club,club,spade}
            &(\varnothing,\{\clubsuit\},\{\clubsuit,\spadesuit\})\\
            \label{eq: spade,club}
            &(\varnothing,\{\spadesuit\},\{\clubsuit\}).
        \end{align}
        Given a state $T\subseteq S$, let $T_V = T \cap V$ and $T_{\n{V}} = T \cap \n{V}$.
        When $x_j \in T_V \Leftrightarrow \n{x_j} \notin T_{\n{V}}$ for every $j$, then $T_V\cup T_{\n{V}}$ encodes an assignment of $\varphi$ in which the variables having true value are those in $T_V$ and the variables having false value are those in $T_{\n{V}}$. 
        In this case we say that $T$ is a \textit{well-formed} state of $\sA$ and the reactions of type (\ref{eq: 0,x_i,neg_x_i}),(\ref{eq: 0,neg_x_i,x_i}) preserve $T_V\cup T_{\n{V}}$, i.e., $T_V\cup T_{\n{V}} \subseteq \res_{\sA}(T)$. 
        Instead, if $T$ is not a well-formed state then we distinguish two cases:
        \begin{itemize}
            \item if $\exists x_i,\n{x_i} \in T_V\cup T_{\n{V}}$ then $x_i,\n{x_i} \notin \res_{\sA}(T)$, since neither $a_i$ nor $\n{a_i}$ is enabled;
             \item if $\exists x_i,\n{x_i} \notin T_V\cup T_{\n{V}}$ then $x_i,\n{x_i} \in \res_{\sA}(T)$, since both $a_i$ and $\n{a_i}$ are enabled.
        \end{itemize}
        In both cases, $T\ne \res_{\sA}(T)$; so if $T$ is a fixed point, $T$ is a well-formed state. 
        For well-formed states, we can also give an interpretation of the reactions of type (\ref{eq: n-neg_pos_spade}): they evaluate each disjunctive clause (which is not satisfied if and only if no positive variables are set to true and no negative ones to false) and generate $\spadesuit$ when $\varphi$ itself is not satisfied by $T_V\cup T_{\n{V}}$. 
        
        Consider the dynamic of the reaction system restricted to a well-formed state $Y\subseteq V \cup \n{V}$. If $Y$ does not satisfy $\varphi$ then there are no fixed points among the well-formed states containing $Y$ because of the following (the arrows represent function $\res_{\sA}$):        
        \begin{center}
            \begin{tikzcd}[row sep=tiny]
             Y \arrow[dr]    &   &\\
                                        & Y \cup \{\clubsuit,\spadesuit\} \arrow[r, bend left] &Y \cup \{\spadesuit\} \arrow[l, bend left] \\
              Y \cup\{\clubsuit\}  \arrow[ru] & &
        \end{tikzcd}
        \end{center}
        Instead, if $Y$ satisfies $\varphi$, then $Y \cup \{\clubsuit\}$ is a fixed point since:
        \begin{center}
            \begin{tikzcd}[row sep=tiny]
              Y \cup\{\spadesuit\}  \arrow[r] & Y \cup\{\clubsuit,\spadesuit\} \arrow[r, bend left] & Y \arrow[l, bend left] &Y \cup\{\clubsuit\}\arrow[loop right]&\quad
        \end{tikzcd}
        \end{center}
        In particular, $\sA$ has a fixed point if and only $\varphi$ is satisfiable. 
        The mapping $\varphi \mapsto \sA$ is computable in polynomial time, hence deciding on the existence of fixed points for reactantless RS is \NP-hard.
    \end{proof}
\end{teor}

As an immediate consequence of Theorem~\ref{teor: reactantless_exist_fix_point} we obtain that deciding if there exists a state that is a common fixed point of two reactantless reaction systems remains \NP-complete.

\begin{cor}
    \label{cor: reactantless_common_fix_point}
    Given $\sA, \sB \in \RS(0,\infty)$ with a common background set $S$, it is \NP-complete to decide whether $\sA$ and $\sB$ have a common fixed point.
    \begin{proof}
        The problem is in \NP (see Table~\ref{tab: problem_logic_class}). 
        By Theorem~\ref{teor: reactantless_exist_fix_point}, when $\sA = \sB$ the problem is \NP-complete.
    \end{proof}
\end{cor}

With a small adaptation of the proof of Theorem~\ref{teor: reactantless_exist_fix_point}, deciding if a fixed point attractor exists is still an \NP-complete problem. 
The \NP-completeness for $\RS(\infty,\infty)$ is proved in \cite[Corollary 3]{formenti2014fixed}; the following theorem proves it for the case of reactantless systems.

\begin{teor}
    \label{teor: reactantless_fix_point_attr}
    Given $\sA=(S,A) \in \RS(0,\infty)$, it is \NP-complete to decide if $\sA$ has a fixed point attractor.
    \begin{proof}
        The problem is in \NP, as highlighted in Table~\ref{tab: problem_logic_class}. 
        In order to show \NP-hardness, we reduce \textsc{sat} \cite{papadimitriou1994computational} to this problem. Given a Boolean formula $\varphi = \varphi_1 \land \cdots \land \varphi_m$ in CNF, we construct the same reaction system $\sA$ of Theorem~\ref{teor: reactantless_exist_fix_point} and we substitute reaction (\ref{eq: club,club,spade}) with $(\varnothing, \{\clubsuit\},\{\clubsuit\})$. In this way, if $Y\subseteq V \cup \n{V}$ is a well-formed state satisfying $\varphi$ we have:
        \begin{center}
            \begin{tikzcd}[row sep=tiny]
            Y \cup\{\clubsuit,\spadesuit\} \arrow[r]  &  Y \arrow[rd] &\\
                                        &  &Y \cup\{\clubsuit\}\arrow[loop right] \\
              &Y \cup\{\spadesuit\}  \arrow[ru] & 
            \end{tikzcd}
        \end{center}
        which means that $Y\cup \{\clubsuit\}$ is a fixed point reachable from $Y$ or $Y\cup \{\spadesuit\}$. 
        In the other cases (either a well-formed state not satisfying $\varphi$ or a not well-formed state), $T\subseteq S$ is never a fixed point, as in the proof of Theorem~\ref{teor: reactantless_exist_fix_point}. 
        Since the mapping  $\varphi \mapsto \sA$ is computable in polynomial time, deciding on the existence of a fixed points attractor for reactantless RS is \NP-hard.
    \end{proof}
\end{teor}

\begin{cor}
    \label{cor: reactantless_common_fix_point_attr}
    Given $\sA, \sB \in \RS(0,\infty)$ with a common background set $S$, it is \NP-complete to decide whether $\sA$ and $\sB$ have a common fixed point attractor.
    \begin{proof}
        The problem is in \NP (see Table~\ref{tab: problem_logic_class}).
        By Theorem~\ref{teor: reactantless_fix_point_attr}, when $\sA = \sB$ the problem is \NP-hard.
    \end{proof}
\end{cor}

In contrast, if we consider the existence of a fixed point not attractor the problem is in \SigtwoP; the following theorem proves the \SigtwoP-hardness.

\begin{teor}
    \label{teor: reactantless_fix_point_not_attr}
    Given $\sA\in \RS(0,\infty)$, it is \SigtwoP-complete to decide if $\sA$ has a fixed point which is not an attractor.
    \begin{proof}
        The problem is in \SigtwoP (see Table~\ref{tab: problem_logic_class}).
        Consider the converse problem, i.e. decide if all fixed points are attractors.
        In order to show \pItwoP-hardness of the latter, we construct a reduction from \textsc{$\forall\exists$sat}~\cite{STOCKMEYER19761}.
        Given a quantified Boolean formula $(\forall V_1)(\exists V_2)\varphi$ over $V = \{x_1,\dots,x_n\}$, with $V_1 \subseteq V$, $V_2 = V \setminus V_1$ and $ \varphi = \varphi_1 \land \cdots \land \varphi_m$ quantifier-free and in CNF, let $\heartsuit_S \coloneqq \{\heartsuit_i: 1\le i\le n\}$ be a set of extra entities that are not contained in $V\cup \n{V} \cup C$ and $V,\n{V}, C$ are as in Definition~\ref{def:sat_entities}.
        We define a RS $\sA$ with background set $S\coloneqq V \cup \n{V} \cup C \cup \heartsuit_S \cup \{\clubsuit,\diamondsuit,\spadesuit\}$ and the reactions
        \begin{align}
            \label{eq: empty,nx+club+spade,varphi}
            &(\varnothing,\{\n{x},\clubsuit,\spadesuit\},\{\varphi_j\})  \:\quad\quad\qquad\qquad\qquad\text{for } \varphi_j \in C &\text{ and } &\n{x} \in \n\posV(\varphi_j) \\
            \label{eq: empty,x+club+spade,varphi}
            &(\varnothing,\{x,\clubsuit,\spadesuit\},\{\varphi_j\})	 \:\quad\quad\qquad\qquad\qquad\text{for } \varphi_j \in C &\text{ and } &x \in \negV(\varphi_j) \\
            \label{eq: empty,nneg+pos+club+spade,spade}
            &(\varnothing,\n\negV(\varphi_j) \cup \posV(\varphi_j)\cup \{\clubsuit,\spadesuit\},\{\spadesuit\}) & \text{ for } &1\le j \le m\\
            \label{eq: empty,x_i+club+spade,heartsuit_i}
            &(\varnothing,\{ x_i,\clubsuit,\spadesuit\},\{\heartsuit_i\}) 		& \text{ for } &1\le i \le n\\
            \label{eq: empty,nx_i+club+spade,heartsuit_i}
            &(\varnothing,\{\n{x_i},\clubsuit,\spadesuit\},\{\heartsuit_i\}) 		& \text{ for } &1\le i \le n\\
            \label{eq: empty,x_i+nx_i+club+spade,spade}
            &(\varnothing,\{ x_i,\n{x_i},\clubsuit,\spadesuit\},\{\spadesuit\}) 		& \text{ for } &1\le i \le n\\
            \label{eq: empty,varphi_j,spade}
            &(\varnothing,\{\varphi_j\},\{\spadesuit\})  				& \text{ for } &1\le j \le m \\
            \label{eq: empty,heartsuit_i,spade}
            &(\varnothing,\{\heartsuit_i\},\{\spadesuit\})  				& \text{ for } &1\le i \le n \\
            \label{eq: empty,nx,x}
            &(\varnothing,\{\n{x}\},\{x\}) 		& \text{ for } &x \in V_1 \\
            \label{eq: empty,x,nx}
            &(\varnothing,\{ x\},\{\n{x}\}) 		& \text{ for } &x \in V_1  \\
            \label{eq: C+heartS+club+V_1+nV_1-fixedpoint}
            &(\varnothing, S\setminus(C\cup \heartsuit_{S}\cup\{\clubsuit\} \cup V_1 \cup \n{V_1}) , C \cup \heartsuit_{S})  	\\
            \label{eq: empty,diamond+spade,club}
            &(\varnothing,\{\diamondsuit,\spadesuit\},\{\clubsuit\})\\
            \label{eq: empty,club+spade,club}
            &(\varnothing,\{\clubsuit,\spadesuit\},\{\clubsuit\})\\
            \label{eq: empty,club+diamond,club+diamond+spade}
            &(\varnothing,\{\clubsuit,\diamondsuit\},\{\clubsuit,\diamondsuit,\spadesuit\}).
        \end{align}
        To study the dynamics of $\sA$ we first analyze the form of its fixed points. 
        \begin{claim}
            \label{claim: T_fixed_point_A_club-diamond-spade}
            If $T$ is a fixed point of $\sA$, then $\clubsuit \in T$, $\diamondsuit \notin T$ and $\spadesuit\notin T$.
            \begin{claimproof}
                If we had $\diamondsuit \in T$, then the only reaction that can produce $\diamondsuit$, i.e., reaction (\ref{eq: empty,club+diamond,club+diamond+spade}), is not enabled, thus $\diamondsuit \notin \res_{\sA}(T) = T$, a contradiction.
                If we had $\spadesuit\in T$ and $\clubsuit\notin T$, then $\{\clubsuit,\diamondsuit\}\subseteq \res_{\sA}(T) = T$ by reaction (\ref{eq: empty,club+diamond,club+diamond+spade}), a contradiction.
                If we had $\spadesuit\in T$ and $\clubsuit\in T$, no reaction that can produce $\clubsuit$ would be enabled (reactions (\ref{eq: empty,diamond+spade,club}), (\ref{eq: empty,club+spade,club}), (\ref{eq: empty,club+diamond,club+diamond+spade})), a contradiction.
                Therefore, it must hold $\diamondsuit \notin T$ and $\spadesuit\notin T$; if in addition we had $\clubsuit \notin T$, then reaction (\ref{eq: empty,club+diamond,club+diamond+spade}) would be enabled and thus  we would have $\{\clubsuit,\diamondsuit,\spadesuit\}\subseteq \res_{\sA}(T) = T$, a contradiction.
                For a visual representation of the dynamics of reactions (\ref{eq: empty,diamond+spade,club}), (\ref{eq: empty,club+spade,club}), (\ref{eq: empty,club+diamond,club+diamond+spade}) see Figure \ref{fig: dynamics_club-diamond-spade}.
                The statement follows.
            \end{claimproof}
        \end{claim}
        \begin{claim}
            \label{claim: classification_fixed_points_A}
            The fixed points of $\sA$ are the states of type 
            \begin{equation}
                \label{eq: T_fixed_point_A_all_attractors}
                 T_U \coloneqq C \cup \heartsuit_S \cup U \cup \n{V_1\setminus U} \cup \{\clubsuit\},
            \end{equation}
            for any $U \subseteq V_1$.
            \begin{claimproof}
                Let $T$ be a fixed point for $\sA$; by Claim ~\ref{claim: T_fixed_point_A_club-diamond-spade} and looking at the products of the reactions it must be $T \subseteq C \cup \heartsuit_S \cup V_1 \cup \n{V_1} \cup \{\clubsuit\}$ with $\clubsuit\in T$.
                Moreover, if we had $ C \cap T \subsetneq C$ then at least one of the reactions of group (\ref{eq: empty,varphi_j,spade}) would be enabled, thus it would be $\spadesuit \in \res_{\sA}(T) = T$, a contradiction.
                If we had $ \heartsuit_S \cap T \subsetneq \heartsuit_S $ then at least one of the reactions of group (\ref{eq: empty,heartsuit_i,spade}) would be enabled, thus we would have $\spadesuit \in \res_{\sA}(T) = T$, a contradiction.
                By reactions (\ref{eq: empty,nx,x}), (\ref{eq: empty,x,nx}) and (\ref{eq: empty,x_i+nx_i+club+spade,spade}), arguing as in the proof of Theorem \ref{teor: reactantless_exist_fix_point}, we get that $T$ must be a well-formed state for $V_1 \cup \n{V_1}$, i.e., $x\in T \cap V_1$ if and only if $\n{x} \notin T \cap \n{V_1}$.
                Therefore, $T$ is of type (\ref{eq: T_fixed_point_A_all_attractors}) with $U = T \cap V_1$.
                Finally, we can immediately check that if a state is of type (\ref{eq: T_fixed_point_A_all_attractors}), then it is a fixed point because of reactions (\ref{eq: empty,nx,x}), (\ref{eq: empty,x,nx}) and (\ref{eq: C+heartS+club+V_1+nV_1-fixedpoint}).
            \end{claimproof}
        \end{claim}
        We are now interested in studying when a fixed point $T_U$, for a given $U\subseteq V_1$, is an attractor.
        Let $T' \ne T_U$ such that $\res_{\sA}(T') = T_U$.
        If it was $C \cap T'\subsetneq C$ then at least one of the reactions of group (\ref{eq: empty,varphi_j,spade}) would be enabled, thus it would be $\spadesuit \in \res_{\sA}(T') = T$, a contradiction; and if it was $\heartsuit_S \cap T' \subsetneq \heartsuit_S$ then at least one of the reactions of group (\ref{eq: empty,heartsuit_i,spade}) would be enabled, thus it would be $\spadesuit \in \res_{\sA}(T') = T$, a contradiction.
        Therefore it must be $C \cup \heartsuit_S \subseteq T'$.
        \begin{claim*}
            If $T' \ne T_U$ and $\res_{\sA}(T') = T_U$ then $\clubsuit \notin T'$, $\diamondsuit \in T'$ and $\spadesuit\notin T'$.
            \begin{claimproof}
                If we had $\spadesuit\in T'$ then reactions (\ref{eq: empty,nx+club+spade,varphi}), (\ref{eq: empty,x+club+spade,varphi}) and (\ref{eq: C+heartS+club+V_1+nV_1-fixedpoint}) would not be enabled, thus we would have $C \cap \res_{\sA}(T') = \varnothing$, a contradiction since $C \subseteq T_U = \res_{\sA}(T')$.
                Therefore we obtain $\spadesuit \notin T'$.
                If we had $\clubsuit \in T'$ we divide two cases:
                \begin{itemize}
                    \item if $\diamondsuit \in T'$ then none of the reactions (\ref{eq: empty,diamond+spade,club}), (\ref{eq: empty,club+spade,club}), (\ref{eq: empty,club+diamond,club+diamond+spade}) would be enabled, therefore we would have $\clubsuit \notin \res_{\sA}(T') = T_U$, a contradiction.
                    \item if $\diamondsuit \notin T'$, the only way to get $C \subseteq \res_{\sA}(T')$ would be with reaction (\ref{eq: C+heartS+club+V_1+nV_1-fixedpoint}), since (\ref{eq: empty,nx+club+spade,varphi}), (\ref{eq: empty,x+club+spade,varphi}) are not enabled if $\clubsuit \in T'$.
                    Therefore to enable (\ref{eq: C+heartS+club+V_1+nV_1-fixedpoint}), we must have $T' \subseteq C\cup \heartsuit_S \cup \{\clubsuit\} \cup V_1 \cup \n{V_1}$, and since $C \cup \heartsuit_S \subseteq T'$, as previously remarked, we obtain $T' = C\cup \heartsuit_{S}\cup\{\clubsuit\} \cup U_1 \cup \n{U_2}$ for some $U_1 \subseteq V_1$ and $U_2 \subseteq V_1$.
                    Since $\res_{\sA}(T')=T_U$ contains a well-formed state for $V_1\cup \n{V_1}$, then $U_1 \cup \n{U_2}$ must be a well-formed state for $V_1\cup \n{V_1}$.
                    In this case we would obtain $\res_{\sA}(T') = C\cup \heartsuit_{S}\cup\{\clubsuit\} \cup U_1 \cup \n{U_2}$, hence we would have $U_1 = U$, $U_2 = V\setminus U$ and thus $T'=T_U$, a contradiction.
                \end{itemize}
                Therefore $\clubsuit \notin T'$.
                Finally, if $\diamondsuit \notin T'$ then $\{\clubsuit,\diamondsuit,\spadesuit\}\subseteq \res_{\sA}(T') = T_U$ (\ref{eq: empty,club+diamond,club+diamond+spade}), a contradiction.
                The statement follows.                
            \end{claimproof}
        \end{claim*}
        By previous claim and remarks, we obtain that if $T'\ne T_U$ is an attractor for $T_U$ then it must be of the form $T' = C \cup \heartsuit_S \cup \{\diamondsuit\} \cup X_1 \cup \n{X_2}$ for some $X_1 \subseteq V$ and $X_2\subseteq V$.
        In this case, reaction (\ref{eq: C+heartS+club+V_1+nV_1-fixedpoint}) is not enabled, thus the fact that $C \cup \heartsuit_S \subseteq \res_{\sA}(T')=T_U$ will only depend on how $X_1 \cup \n{X_2}$ is constructed.
        We consider two cases:
        \begin{itemize}
            \item if $\exists x_i,\n{x_i} \in X_1 \cup \n{X_2}$ then $\heartsuit_i \notin \res_{\sA}(T')$, since neither the $i$-th reaction of (\ref{eq: empty,x_i+club+spade,heartsuit_i}) nor the $i$-th reaction of (\ref{eq: empty,nx_i+club+spade,heartsuit_i}) are enabled;
             \item if $\exists x_i,\n{x_i} \notin X_1 \cup \n{X_2}$ then $\spadesuit \in \res_{\sA}(T')$, since the $i$-th reaction in (\ref{eq: empty,x_i+nx_i+club+spade,spade}) is enabled.
        \end{itemize}
        Therefore $X_1 \cup \n{X_2}$ must be a well-formed state for $V \cup \n{V}$, and thus $X_2 = V \setminus X_1$.
        We can interpret $X_1 \cup \n{V \setminus X_1}$ as an assignment for $\varphi$, as in the proof of Theorem \ref{teor: reactantless_exist_fix_point}.
        If we had $X_1 \cup \n{V \setminus X_1} \nvDash \varphi$, then one of the reactions (\ref{eq: empty,nneg+pos+club+spade,spade}) would be enabled, thus we would have $\spadesuit \in \res_{\sA}(T')$, a contradiction. 
        Therefore it must hold $X_1 \cup \n{V \setminus X_1} \vDash \varphi$, thus all clauses are satisfied, i.e., each $\varphi_j \in C$ is generated by at least one reaction of type (\ref{eq: empty,club+spade,club}) or (\ref{eq: empty,club+diamond,club+diamond+spade}).
        We obtain
        \begin{equation*}
            \res_{\sA}(T') = C \cup \heartsuit_S \cup \{\clubsuit\} \cup \{x : x \in X_1 \cap V_1\} \cup \{\n{x}: \n{x} \in \n{X_2}\cap \n{V_1}\} = T_U
        \end{equation*}
        thus $U = X_1 \cap V_1$, and therefore if $T_U$ is an attractor there exists an assignment for $\varphi$ such that all variables in $U$ are set to true. 
        The converse is also true, i.e., if there exists an assignment $X \cup \n{V\setminus X}$ such that $U \subseteq X$ and $\varphi$ is satisfied then $T_U$ is reached by $C \cup \heartsuit_S \cup \{\diamondsuit\} \cup X \cup \n{V\setminus X}$.
        Finally, all the fixed points of $\sA$ are attractors if and only if $(\forall V_1) (\exists V_2) \varphi$ is valid.
        
        The mapping $\varphi \mapsto \sA$ is computable in polynomial time, hence deciding if all the fixed points of $\sA$ are attractors is \pItwoP-hard.
        Therefore, deciding if $\sA$ has a fixed point which is not attractor is \SigtwoP-hard.
    \end{proof}
\end{teor}

\begin{cor}
    \label{cor: reactantless_common_fix_not_attr}
    Given $\sA, \sB \in \RS(0,\infty)$ with a common background set $S$, it is \SigtwoP-complete to decide whether $\sA$ and $\sB$ have a common fixed point which is not an attractor.
    \begin{proof}
       The problem is in \SigtwoP (see Table~\ref{tab: problem_logic_class}); by Theorem~\ref{teor: reactantless_fix_point_not_attr}, when $\sA = \sB$ the problem is \SigtwoP-complete.
    \end{proof}
\end{cor}

Since the problems are \SigtwoP-complete for $\RS(0,\infty)$, we have that they are also \SigtwoP-complete for $\RS(\infty,\infty)$, as stated by the following corollaries.

\begin{cor}
    \label{cor: fix_point_not_attr}
    Given $\sA = (S,A)\in \RS(\infty,\infty)$, it is \SigtwoP-complete to decide if $\sA$ has a fixed point which is not an attractor.
\end{cor}

\begin{cor}
    \label{cor: common_fix_not_attr}
    Given $\sA, \sB \in \RS(\infty,\infty)$ with a common background set $S$, it is \SigtwoP-complete to decide whether $\sA$ and $\sB$ have a common fixed point which is not an attractor.
\end{cor}

We now study the problem of deciding if two reaction systems share all their fixed point attractors.

\begin{cor}
    \label{cor: reactantless_all_fix_point_attr}
    Given $\sA, \sB \in \RS(0,\infty)$ with a common background set $S$, it is \pItwoP-complete to decide whether $\sA$ and $\sB$ share all their fixed point attractors.
    \begin{proof}
        The problem is in \pItwoP (see Table~\ref{tab: problem_logic_class}). 
	In order to show \pItwoP-hardness, we reduce \textsc{$\forall\exists$sat} \cite{STOCKMEYER19761} to this problem.
        Consider the reaction system $\sA$ in the proof of Theorem \ref{teor: reactantless_fix_point_not_attr}; we just need to construct a RS $\sB$ over the same background set as $\sA$ such that all the fixed points attractors of $\sB$ are of the form $T_U$ for any $U\subseteq V_1$.
        Therefore we define the reactions of $\sB$ to be the reactions (\ref{eq: empty,varphi_j,spade}), (\ref{eq: empty,heartsuit_i,spade}), (\ref{eq: empty,nx,x}), (\ref{eq: empty,x,nx}), (\ref{eq: C+heartS+club+V_1+nV_1-fixedpoint}), (\ref{eq: empty,diamond+spade,club}), (\ref{eq: empty,club+spade,club}) and (\ref{eq: empty,club+diamond,club+diamond+spade}) from the proof of Theorem \ref{teor: reactantless_fix_point_not_attr}.
        We remark that Claims \ref{claim: T_fixed_point_A_club-diamond-spade} and \ref{claim: classification_fixed_points_A} can be applied to $\sB$, therefore all the fixed points of $\sB$ are the states $T_U$ for any $U\subseteq V_1$.
        Furthermore, all the fixed points of $\sB$ are attractors and coincide with the ones of $\sA$.
        The mapping $\varphi \mapsto (\sA, \sB)$ is computable in polynomial time, hence deciding if two RS share all their fixed point attractors is \pItwoP-hard.
    \end{proof}
\end{cor}

We now study the problem of deciding whether two reaction systems share all their fixed points.
The problem is \coNP-complete for $\RS(\infty,\infty)$ \cite[Theorem 3]{formenti2014fixed}, and this is also true for reactantless reaction systems, as proved in the following theorem.

\begin{teor}
    \label{teor: reactantless_all_fix_point}
    Given $\sA, \sB \in \RS(0,\infty)$ with a common background set $S$, it is \coNP-complete to decide whether $\sA$ and $\sB$ share all their fixed points.
    \begin{proof}
        The problem lies in \coNP (see Table~\ref{tab: problem_logic_class}).
        In order to show \coNP-completeness, we reduce \textsc{validity} \cite{papadimitriou1994computational} to this problem. Given a Boolean formula $\varphi = \varphi_1 \lor \cdots \lor \varphi_m$ in DNF over the variables $V = \{x_1,\dots,x_n\}$, we define a RS $\sA$ with background set $S\coloneqq V \cup \n{V} \cup \{\heartsuit\} \cup \{\clubsuit\}$ (where the sets are as in Definition~\ref{def:sat_entities}) and the reactions:
        \begin{align}
            \label{eq: neg_n-pos_heart}
            &(\varnothing, \negV(\varphi_j) \cup \n\posV(\varphi_j) \cup \{\clubsuit\}, \{\heartsuit\}) & \text{ for } &1\le j \le m\\
            &(\varnothing, \{x_i\}, \{\n{x_i}\}) & \text{ for } &1\le i \le n\\
            &(\varnothing,\{\n{x_i}\}, \{x_i\}) & \text{ for } &1\le i \le n\\
            \label{eq: heart,heart}
            &(\varnothing,\{\heartsuit\},\{\heartsuit,\clubsuit\}).
        \end{align}
        As in the proof of Theorem~\ref{teor: reactantless_exist_fix_point}, if $T$ is not a well-formed state then $T$ is not a fixed point. 
        Reactions of type (\ref{eq: neg_n-pos_heart}) evaluate each conjunctive clause (which is satisfied if and only if no positive variables are set to false and no negative ones to true) and generate $\heartsuit$ when $\varphi$ itself is satisfied by $T \cap (V \cup \n{V})$.
        Consider the dynamic of the reaction system restricted to a well-formed state $Y \subseteq V \cup \n{V}$. 
        If $Y$ does not satisfy $\varphi$ there are no fixed points since:
        \begin{center}
            \begin{tikzcd}[row sep=tiny]
              Y \cup\{\heartsuit\}  \arrow[r] & Y \arrow[r, bend left] & Y \cup\{\heartsuit,\clubsuit\}\arrow[l, bend left] &Y \cup\{\clubsuit\}\arrow[l]
        \end{tikzcd}
        \end{center}
        where the arrow represent the function $\res_{\sA}$.
        Instead, if $Y$ satisfies $\varphi$, $Y \cup \{\heartsuit\}$ is a fixed point since:
        \begin{center}
            \begin{tikzcd}[row sep=tiny]
              Y \cup\{\heartsuit\}  \arrow[loop left] & Y \arrow[r, bend left] & Y \cup\{\heartsuit,\clubsuit\}\arrow[l, bend left] &Y \cup\{\clubsuit\}.\arrow[l]
            \end{tikzcd}
        \end{center}
        Finally, the fixed points of $\sA$ are the well-formed states $Y \cup \{\heartsuit\}$ such that $Y \vDash \varphi$.
        Now, let $\sB$ be defined by the following reactions:
        \begin{align*}
            &(\varnothing, \{x_i\}, \{\n{x_i}\}) & \text{ for } &1\le i \le n\\
            &(\varnothing,\{\n{x_i}\}, \{x_i\}) & \text{ for } &1\le i \le n\\
            &(\varnothing,\{\clubsuit\},\{\heartsuit\})\\
            &(\varnothing,\{\heartsuit\},\{\heartsuit,\clubsuit\}).
        \end{align*}
        In a similar way as above, the fixed points of $\sB$ are the states $Y \cup \{\heartsuit\}$ where $Y\subseteq V\cup \n{V}$ is well-formed.
        We can conclude that $\sA$ and $\sB$ share all fixed points exactly when all assignments satisfy $\varphi$, i.e., $\varphi$ is a tautology. 
        Since the mapping $\varphi \mapsto (\sA,\sB)$ is computable in polynomial time, the problem is \coNP-hard.
    \end{proof}
\end{teor}

Note that in the proof of Theorem~\ref{teor: reactantless_all_fix_point}, the fixed points of $\sA$ and $\sB$ are not attractors: this implies the following result.
\begin{cor}
    \label{cor: reactantless_all_fix_pointge}
    Given $\sA, \sB \in \RS(0,\infty)$ with a common background set $S$, it is \coNP-complete to decide whether $\sA$ and $\sB$ share all their fixed points which are not attractors. 
\end{cor}
Since the problem is \coNP-complete for $\RS(0,\infty)$, we have that it is also \coNP-complete for $\RS(\infty,\infty)$, as stated by the following corollary.
\begin{cor}
    \label{cor: all_fix_pointge}
    Given $\sA, \sB \in \RS(\infty,\infty)$ with a common background set $S$, it is \coNP-complete to decide whether $\sA$ and $\sB$ share all their fixed points which are not attractors.
\end{cor}

\section{Fixed Points for Inhibitorless RS}
\label{sec: fix_point_infty_0}

In this section, we prove \NP-hardness and \coNP-hardness for problems of fixed points in the class of inhibitorless RS.

The problem of deciding the existence of a fixed point is entirely trivial for $\RS(\infty,0)$ thanks to the Knaster-Tarski theorem, as first remarked in \cite{manzoni2014simple}.
On the contrary, the following theorem shows that it is \NP-complete to decide whether two inhibitorless RS have a common fixed point.

\begin{teor}
    \label{teor: inhibitorless_common_fix_point}
    Given $\sA, \sB \in \RS(\infty,0)$ with a common background set $S$, it is \NP-complete to decide whether $\sA$ and $\sB$ have a common fixed point.
    \begin{proof}
        The problem is in \NP (see Table~\ref{tab: problem_logic_class}). 
        In order to show \NP-hardness, we reduce \textsc{sat} \cite{papadimitriou1994computational} to this problem. Given a Boolean formula $\varphi = \varphi_1 \land \cdots \land \varphi_m$ in CNF over the variables $V = \{x_1,\dots,x_n\}$, let $\heartsuit_S \coloneqq \{\heartsuit_i: 1\le i\le n\}$ be a set of extra entities that are not contained in $V\cup \n{V}$.
        We define a RS $\sA$ with background set $S\coloneqq V \cup \n{V} \cup \heartsuit_S \cup \{\spadesuit\}$ (with $\spadesuit\not\in V \cup \n{V} \cup \heartsuit_S$ and $V,\n{V}$ as in Definition~\ref{def:sat_entities}) and the reactions
        \begin{align}
            \label{eq: neg_n-pos-heart_spade}
            &(\negV(\varphi_j) \cup \n\posV(\varphi_j) \cup \heartsuit_S, \varnothing, \{\spadesuit\}) & \text{ for } &1\le j \le m\\
            \label{eq: x_heart_S,,heart_i_x}
            &(\{x_i\} \cup \heartsuit_S,\varnothing, \{\heartsuit_i,x_i\}) & \text{ for } &1\le i \le n\\
            \label{eq: eq: negx_heart_S,,heart_i_negx}
            &(\{\n{x_i}\} \cup \heartsuit_S,\varnothing, \{\heartsuit_i,\n{x_i}\}) & \text{ for } &1\le i \le n\\
            \label{eq: eq: x_negx,,spade}
            &( \{x_i,\n{x_i}\} \cup \heartsuit_S,\varnothing, \{\spadesuit\}) & \text{ for } &1\le i \le n\\
            \label{eq: spade,,spade}
            &(\{\spadesuit\}\cup \heartsuit_S,\varnothing,\{\spadesuit\}).
        \end{align}
        Note that for all $Y\subseteq  V \cup \n{V}$, and for every $Z_{\heartsuit} \subsetneq \heartsuit_{S}$ it holds
        \begin{equation}
            \label{eq: res_(Y + Z_heart)=empty}
            \res_{\sA}(Y \cup Z_{\heartsuit}) = \res_{\sA}(Y \cup Z_{\heartsuit}\cup \{\spadesuit\}) = \varnothing = \res_{\sA}(\varnothing)
        \end{equation}
        because no reaction is enabled. We thus consider states $T\subseteq S$ such that $\heartsuit_{S} \subseteq T$.
        For every $Y \subseteq V \cup \n{V}$, we define $\heartsuit_Y\coloneqq \{\heartsuit_i: x_i \in Y \lor \n{x_i}\in Y\} \subseteq \heartsuit_S$. 
        Note that $\heartsuit_Y = \res_{\sA}(Y\cup \heartsuit_S) \cap \heartsuit_S = \res_{\sA}(Y\cup \heartsuit_S\cup \{\spadesuit\})\cap \heartsuit_S$, so when $\heartsuit_Y \subsetneq \heartsuit_S$, the states $Y\cup \heartsuit_S$ and $Y\cup \heartsuit_S\cup \{\spadesuit\}$ reach $\varnothing$ in two steps. 
        In particular, if $T\ne \varnothing$ is a fixed point then it must be of the form $T=Y\cup \heartsuit_Y$ or $T=Y\cup \heartsuit_Y \cup \{\spadesuit\}$ with $Y \subseteq V \cup \n{V}$ and $\heartsuit_Y = \heartsuit_S$.
        We remark that $\heartsuit_Y = \heartsuit_S$ means that $x_i \in Y$ or $\n{x_i}\in Y$ for all $1\le i \le n$.
        We divide two cases:
        \begin{enumerate}[(i)]
            \item $Y$ is not a well-formed state. 
            Since $\heartsuit_{Y} = \heartsuit_{S}$, there exist $x_i,\n{x_i}\in Y$, so $\spadesuit$ is generated by one of the reactions of type (\ref{eq: eq: x_negx,,spade}). 
            We obtain that $Y\cup \heartsuit_S \cup \{\spadesuit\}$ is a fixed point reachable from $Y\cup \heartsuit_S$.
            \item $Y$ is a well-formed state. 
            If $Y\vDash \varphi$ then no reaction of type (\ref{eq: neg_n-pos-heart_spade}) is enabled, so $Y \cup \heartsuit_S$ is a fixed point (not reachable from any other state). 
            Also in this case, $Y \cup \heartsuit_S\cup \{\spadesuit\}$ is a fixed point, thanks to reaction (\ref{eq: spade,,spade}).
            On the other hand, if $Y\nvDash \varphi$ then $Y \cup \heartsuit_S\cup \{\spadesuit\}$ is a fixed point reachable from $Y \cup \heartsuit_S$.
        \end{enumerate}
        Now, consider the RS $\sB$ given by the following reactions:
        \begin{align}
            &(\varnothing, \varnothing, \heartsuit_S) \\
            &(\{x_i\} ,\varnothing, \{x_i\}) & \text{ for } &1\le i \le n\\
            &(\{\n{x_i}\},\varnothing, \{\n{x_i}\})& \text{ for } &1\le i \le n.
        \end{align}
        The fixed points of $\sB$ are the states $Y \cup \heartsuit_S$ for all $Y\subseteq V \cup \n{V}$. 
        We can conclude that $\sA$ and $\sB$ share a fixed point exactly when there exists an assignment satisfying $\varphi$, i.e., $\varphi$ is satisfiable. 
        Since the mapping $\varphi \mapsto (\sA,\sB)$ is computable in polynomial time, the problem is \NP-hard.
    \end{proof}
\end{teor}

We next show that determining whether two inhibitorless RS have a common fixed point attractor is \NP-complete. The proof is an adaptation of the proof of Theorem~\ref{teor: inhibitorless_common_fix_point}.

\begin{cor}
    \label{cor: inhibitorless_common_fix_point_attr}
    Given $\sA, \sB \in \RS(\infty,0)$ with a common background set $S$, it is \NP-complete to decide whether $\sA$ and $\sB$ have a common fixed point attractor.
    \begin{proof}
        The problem is in \NP (see Table~\ref{tab: problem_logic_class}).
        Following the proof of Theorem~\ref{teor: inhibitorless_common_fix_point}, we just need to ensure that the fixed points of $\sA$ are attractors, so we delete reaction (\ref{eq: spade,,spade}) from the reactions of $\sA$. In this way, when $Y\subset V \cup \n{V}$ and $Y \vDash \varphi$, we have that
        \begin{equation*}
            \res_{\sA}(Y \cup \heartsuit_S\cup \{\spadesuit\}) =  Y \cup \heartsuit_S = \res_{\sA}(Y \cup \heartsuit_S),
        \end{equation*}
        thus $Y \cup \heartsuit_S$ is a fixed point reachable from $Y \cup \heartsuit_S\cup \{\spadesuit\}$.
        We can conclude that $\sA$ and $\sB$ share a fixed point attractor exactly when there exists an assignment satisfying $\varphi$, i.e., $\varphi$ is satisfiable.
        Since the mapping $\varphi \mapsto (\sA,\sB)$ is computable in polynomial time, the problem is \NP-hard.\qedhere
    \end{proof}
\end{cor}

We now consider the problem of deciding if two RS share all their fixed points and prove that, like in the case of reactantless systems, this problem is \coNP-complete.

\begin{teor}
    \label{teor: inhibitorless_all_fix_point}
    Given $\sA, \sB \in \RS(\infty,0)$ with a common background set $S$, it is \coNP-complete to decide whether $\sA$ and $\sB$ share all their fixed points.
    \begin{proof}
        The problem lies in \coNP (see Table~\ref{tab: problem_logic_class}). 
        In order to show \coNP-completeness, we reduce \textsc{validity} \cite{papadimitriou1994computational} to this problem. 
        Given a Boolean formula $\varphi = \varphi_1 \lor \cdots \lor \varphi_m$ in DNF over the variables $V = \{x_1,\dots,x_n\}$,  let $\heartsuit_S \coloneqq \{\heartsuit_i: 1\le i\le n\}$ be a set of extra entities. 
        We define a RS $\sA$ with background set $S\coloneqq V \cup \n{V} \cup \heartsuit_S \cup \{\heartsuit\}$ (with $\heartsuit\notin V \cup \n{V} \cup \heartsuit_S$ and $V,\n{V}$ as in Definition~\ref{def:sat_entities}) and the reactions
        \begin{align}
            \label{eq: n-neg_pos-hearts_heart}
            &(\n\negV(\varphi_j) \cup \posV(\varphi_j) \cup \heartsuit_S \cup \{\heartsuit\}, \varnothing, \{\heartsuit\}) & \text{ for } &1\le j \le m\\
            &(\{x_i\} \cup \heartsuit_S,\varnothing, \{\heartsuit_i,x_i\}) & \text{ for } &1\le i \le n\\
            &(\{\n{x_i}\} \cup \heartsuit_S,\varnothing, \{\heartsuit_i,\n{x_i}\}) & \text{ for } &1\le i \le n\\
            \label{eq: eq: x_negx,,heart}
            &( \{x_i,\n{x_i}\} \cup \heartsuit_S,\varnothing, \{\heartsuit\}) & \text{ for } &1\le i \le n.
        \end{align}
         For every $Y \subseteq V \cup \n{V}$, we define $\heartsuit_Y\coloneqq \{\heartsuit_i: x_i \in Y \lor \n{x_i}\in Y\} \subseteq \heartsuit_S$. As in the proof of Theorem~\ref{teor: inhibitorless_common_fix_point}, if $T\ne \varnothing$ is a fixed point then it must be of the form $T=Y\cup \heartsuit_Y$ or $T=Y\cup \heartsuit_Y \cup \{\heartsuit\}$ with $\heartsuit_Y = \heartsuit_S$. 
         We divide two cases:
        \begin{enumerate}[(i)]
            \item $Y$ is not a well-formed state. Then, since $\heartsuit_{Y} = \heartsuit_{S}$, there exist $x_i,\n{x_i}\in Y$, so $\heartsuit$ is generated by one of the reactions of type (\ref{eq: eq: x_negx,,heart}). We get that $Y\cup \heartsuit_Y \cup \{\heartsuit\}$ is a fixed point reachable from $Y\cup \heartsuit_Y$.
            \item $Y$ is a well-formed state. Then, if $Y\vDash \varphi$, a reaction of type (\ref{eq: n-neg_pos-hearts_heart}) is enabled by $Y\cup \heartsuit_Y \cup \{\heartsuit\}$, so $Y\cup \heartsuit_Y \cup \{\heartsuit\}$ is a fixed point (not reachable from any other state). 
            In this case, also $Y \cup \heartsuit_S$ is a fixed point since reactions of type (\ref{eq: n-neg_pos-hearts_heart}) are not enabled.
            On the other hand, if $Y\nvDash \varphi$ then $Y \cup \heartsuit_S$ is a fixed point reachable from $Y \cup \heartsuit_S \cup \{\heartsuit\}$.
        \end{enumerate}
        Now, consider the RS $\sB$ given by the following reactions:
        \begin{align}
            \label{eq: hearts,,heart}
            &(\{\heartsuit\} \cup \heartsuit_S, \varnothing, \{\heartsuit\}) \\
            &(\{x_i\} \cup \heartsuit_S,\varnothing, \{\heartsuit_i,x_i\}) & \text{ for } &1\le i \le n\\
            &(\{\n{x_i}\} \cup \heartsuit_S,\varnothing, \{\heartsuit_i,\n{x_i}\}) & \text{ for } &1\le i \le n\\
            &( \{x_i,\n{x_i}\} \cup \heartsuit_S,\varnothing, \{\heartsuit\}) & \text{ for } &1\le i \le n.
        \end{align}
        With a similar analysis as above, for every well-formed state $Y \subseteq V \cup \n{V}$ the states $Y \cup \heartsuit_S$, $Y\cup \heartsuit_Y \cup \{\heartsuit\}$ are fixed points (not attractors), and for every not-well-formed state $Y \subseteq V \cup \n{V}$ such that $\heartsuit_Y=\heartsuit_S$ the state $Y\cup \heartsuit_Y \cup \{\heartsuit\}$ is a fixed point reachable from $Y \cup \heartsuit_S$.
        We can conclude that $\sA$ and $\sB$ share all fixed points exactly when all assignments satisfy $\varphi$, i.e., $\varphi$ is a tautology. 
        Since the mapping $\varphi \mapsto (\sA,\sB)$ is computable in polynomial time, the problem is \coNP-hard.\qedhere  
    \end{proof}
\end{teor}

\begin{cor}
    \label{cor: inhibitorless_all_fix_pointge}
    Given $\sA, \sB \in \RS(\infty,0)$ with a common background set $S$, it is \coNP-complete to decide whether $\sA$ and $\sB$ share all their fixed points that are not attractors.
    \begin{proof}
        The problem is in \coNP (see Table~\ref{tab: problem_logic_class}).
        The \coNP-hardness follows from the same construction of Theorem~\ref{teor: inhibitorless_all_fix_point}.
    \end{proof}
\end{cor}

In contrast, deciding whether two inhibitorless reaction systems share all their fixed points that are attractors is \pItwoP-complete.

\begin{teor}
    \label{teor: inhibitorless_all_fix_point_attr}
    Given $\sA, \sB \in \RS(\infty,0)$ with a common background set $S$, it is  \pItwoP-complete to decide if $\sA$ and $\sB$ share all their fixed point attractors.
    \begin{proof}
	The problem is in \pItwoP (see Table~\ref{tab: problem_logic_class}). 
	In order to show \pItwoP-hardness, we reduce \textsc{$\forall\exists$sat} \cite{STOCKMEYER19761} to this problem.
	Given a quantified Boolean formula $(\forall V_1)(\exists V_2)\varphi$ over $V = \{x_1,\dots,x_n\}$, with $V_1 \subseteq V$, $V_2 = V \setminus V_1$ and $ \varphi = \varphi_1 \land \cdots \land \varphi_m$ quantifier-free and in CNF, let $V'_1 = \{x' : x \in V_1\}$ and $\heartsuit_S \coloneqq \{\heartsuit_i: 1\le i\le n\}$ be extra sets of entities.
	We define $\sA$ a RS with background set $S\coloneqq V \cup \n{V} \cup V'_1 \cup C \cup \heartsuit_S \cup \{\clubsuit,\spadesuit\}$ (see also Definition~\ref{def:sat_entities}) and the reactions
	\begin{align}
            &(\{x\},\varnothing,\{\varphi_j\})  			& \text{ for } &1\le j \le m \text{ and } x \in \posV(\varphi_j)\\
            &(\{\n{x}\},\varnothing,\{\varphi_j\})	& \text{ for } &1\le j \le m \text{ and } \n{x} \in \n\negV(\varphi_j)\\
                \label{eq: varphi_j_not_satisfied}
            &(\negV(\varphi_j) \cup \n\posV(\varphi_j),\varnothing,\{\spadesuit\}) & \text{ for } &1\le j \le m\\
            &(\{ x_i\},\varnothing,\{\heartsuit_i\}) 		& \text{ for } &1\le i \le n\\
            &(\{\n{x_i}\},\varnothing,\{\heartsuit_i\}) 		& \text{ for } &1\le i \le n\\
            &(\{ x_i,\n{x_i}\},\varnothing,\{\clubsuit\}) 		& \text{ for } &1\le i \le n\\
            &(\{ x\},\varnothing,\{x'\}) 		& \text{ for } &x \in V_1\\
            \label{eq: C_heart_S_fixed_point}
            &(C\cup \heartsuit_{S},\varnothing, C \cup \heartsuit_{S}) 		\\
            \label{eq: C_heart_S_x'_fixed_point}
            &(C\cup \heartsuit_{S} \cup \{x'\},\varnothing,\{x'\}) 		& \text{ for }& x' \in V'_1\\
            \label{eq: C_heart_s_x_i,empty,club}
            &(C\cup \heartsuit_{S} \cup \{x_i\},\varnothing,\{\clubsuit\}) 		& \text{ for } &1\le i \le n\\
            \label{eq: C_heart_s_negx_i,empty,club}
            &(C\cup \heartsuit_{S} \cup \{\n{x_i}\},\varnothing,\{\clubsuit\}) 		& \text{ for } &1\le i \le n\\
            \label{eq: club,empty,club}
            &(\{\clubsuit\},\varnothing,\{\clubsuit\})\\
            \label{eq: spade,empty,club}
            &(\{\spadesuit\},\varnothing,\{\clubsuit\}).
        \end{align}
	We first note that for any $T \subseteq S\setminus \{\clubsuit\}$ we have $\res_{\sA}(T \cup \{\clubsuit\}) = \res_{\sA}(T) \cup \{\clubsuit\}$.
	We start by determining the fixed points of $\sA$.
	We notice that if $T$ is a fixed point then it must be $T = \res_{\sA}(T)\subseteq C \cup \heartsuit_{S} \cup V_1'\cup \{\clubsuit,\spadesuit\}$, because this is the union of the products of all reactions.
	Furthermore, the only way for $\spadesuit$ to be part of the product is through reactions (\ref{eq: varphi_j_not_satisfied}) which use reactants in $V \cup \n{V}$ that are never part of a fixed point, as we already noticed.
  
	For the same reason, the only reactions that can give rise to a fixed point are (\ref{eq: C_heart_S_fixed_point}), (\ref{eq: C_heart_S_x'_fixed_point}), (\ref{eq: club,empty,club}) and thus we deduce that all the fixed points of $\sA$ are $\varnothing$, $\{\clubsuit\}$, and those of the form $C\cup \heartsuit_{S}\cup U \cup \{\clubsuit\}$ or $C\cup \heartsuit_{S}\cup U$, with $U \subseteq V'_1$.
	The states $\varnothing$, $\{\clubsuit\}$ and $C\cup \heartsuit_{S}\cup U \cup \{\clubsuit\}$ are all attractors, so we now focus on understanding when a state of the form $C\cup \heartsuit_{S}\cup U$ is a fixed point attractor for a given $U \subseteq V'_1$.
		
        Let $T\ne C\cup \heartsuit_{S}\cup U$ be a state such that $\res_{\sA}(T) = C\cup \heartsuit_{S}\cup U$.
	Since $\clubsuit \notin \res_{\sA}(T)$, neither (\ref{eq: club,empty,club}) or (\ref{eq: spade,empty,club}) can be enabled, thus $\clubsuit \notin T$ and $\spadesuit \notin T$.
	We obtain that $T$ is of the form
	\begin{equation*}
		T = T_V \cup T_{\n{V}} \cup T_{V'_1} \cup T_C \cup T_{\heartsuit_{S}}
	\end{equation*}
	where $T_V \coloneqq T \cap V$ and analogously for $T_{\n{V}}$, $T_{V'_1}$, $T_C$ and $T_{\heartsuit_{S}}$.
	If we had $T_C = C$ and $T_{\heartsuit_{S}} = \heartsuit_{S}$, then we would also have $T_V = T_{\n{V}} = \varnothing$ as otherwise $\clubsuit \in \res_{\sA}(T)$ would be generated by at least one of reactions of type (\ref{eq: C_heart_s_x_i,empty,club}) or (\ref{eq: C_heart_s_negx_i,empty,club}).
	Thus, in this case, we would have $C \cup \heartsuit_{S} \cup U = \res_{\sA}(T) = C \cup \heartsuit_{S} \cup T_{V'_1} = T$, which is a contradiction.
	Therefore it must hold $T_C\subsetneq C$ and $T_{\heartsuit_{S}} \subsetneq \heartsuit_{S}$, and, collecting all together, the reactions of type (\ref{eq: C_heart_S_fixed_point}), (\ref{eq: C_heart_S_x'_fixed_point}), (\ref{eq: C_heart_S_x'_fixed_point}), (\ref{eq: C_heart_s_x_i,empty,club}), (\ref{eq: C_heart_s_negx_i,empty,club}), (\ref{eq: club,empty,club}) and (\ref{eq: spade,empty,club}) are not enabled.
  
	We further remark that as in the previous proofs, $T_V \cup T_{\n{V}}$ must be a well-formed assignment for $\varphi$.
	Furthermore if $T_V \cup T_{\n{V}} \vDash \varphi$ then $\res_{\sA}(T) = C \cup \heartsuit_S \cup \{x' : x \in T_V \cap V_1\}$.
	We can also remark that $\spadesuit \in \res_{\sA}(T)$ if and only if there exists a $\varphi_j \in C$ that is not satisfied by the assignment corresponding to $T_V \cup T_{\n{V}}$.
	We deduce that $C \cup \heartsuit_S \cup U$ is a fixed point attractor if and only if there exists an assignment $X \cup \n{V\setminus X}$ such that $\{x : x' \in U\} \subseteq X$ and $X \cup \n{V\setminus X} \vDash \varphi$.
	We conclude that all fixed points of $\sA$ are attractors if and only $(\forall V_1)(\exists V_2)\varphi$ is valid.
  
	We now finally consider a RS $\sB$ with background set $S$ and reactions (\ref{eq: C_heart_S_fixed_point}), (\ref{eq: C_heart_S_x'_fixed_point}), (\ref{eq: club,empty,club}).
	All the fixed points of $\sB$ are attractors and coincide with the ones of $\sA$.
	The mapping $\varphi \mapsto (\sA, \sB)$ is computable in polynomial time, hence deciding if two RS share all their fixed point attractors is \pItwoP-hard.
    \end{proof}
\end{teor}

\begin{cor}
    \label{cor: inhibitorless_fix_point_not_attr}
    Given $\sA \in \RS(\infty,0)$ it is \SigtwoP-complete to decide whether $\sA$ has a fixed point which is not an attractor.
    \begin{proof}
	The problem is in \SigtwoP (see Table~\ref{tab: problem_logic_class}).
	Consider the converse problem, i.e., deciding if all fixed points are attractors.
	The \pItwoP-hardness of the latter follows from the construction of the RS $\sA$ in the proof of Theorem \ref{teor: inhibitorless_all_fix_point_attr}.
	Therefore our problem is \SigtwoP-complete.
    \end{proof}
\end{cor}

\begin{cor}
    \label{cor: inhibitorless_common_fix_point_not_attr}
    Given $\sA, \sB \in \RS(\infty,0)$ with a common background set $S$, it is \SigtwoP-complete to decide whether $\sA$ and $\sB$ have a common fixed point which is not an attractor.
    \begin{proof}
	The problem is in \SigtwoP (see Table~\ref{tab: problem_logic_class}); by Corollary~\ref{cor: inhibitorless_fix_point_not_attr}, when $\sA = \sB$ the problem is \SigtwoP-complete.
    \end{proof}
\end{cor}

\section{Equal Result function}\label{sec:resultFunction}
In this section, we study the problem of deciding if two RS have the same result function.
This problem lies in \coNP and is complete in general RS.
\begin{teor}
    \label{teor: res_A=res_B}
    Given $\sA, \sB \in \RS(\infty,\infty)$ with the same background set $S$, it is \coNP-complete to decide whether $\res_{\sA} = \res_{\sB}$.
    \begin{proof}
        The problem lies in \coNP (see Table~\ref{tab: problem_logic_class}).
        In order to show \coNP-completeness, we reduce \textsc{validity} \cite{papadimitriou1994computational} to this problem. Given a Boolean formula $\varphi = \varphi_1 \lor \cdots \lor \varphi_m$ in DNF over the variables $V = \{x_1,\dots,x_n\}$, build the RS $\sA$ consisting of the background set $S \coloneqq V \cup \{\heartsuit\} $ (with $\heartsuit\notin V$, and $\posV(\varphi_r) \subseteq V$ and $\negV(\varphi_r) \subseteq V$ the set of variables that occur non-negated and negated in $\varphi_r$, respectively) and the following reactions:
        \begin{equation}
            \label{eq: res_A=res_B_A}
            (\posV(\varphi_j),\negV(\varphi_j),\{\heartsuit\}) \qquad \text{for } 1\le j\le m.
        \end{equation}
        For any state $T\subseteq S$, $T \cap V$ encodes a truth assignment of $\varphi$. In this way, the reactions of type (\ref{eq: res_A=res_B_A}) evaluate each conjunctive clause and produce the element $\heartsuit$ if the clause (and thus the whole formula $\varphi$) is satisfied.
        Then the RS behaves as follows:
        \begin{equation*}
            \res_{\sA} (T) = 
            \begin{cases}
            \heartsuit      &\text{if } T \cap V \vDash \varphi\\
            \varnothing     &\text{if } T \cap V \nvDash \varphi.
            \end{cases}
        \end{equation*}
        Now let $\sB$ be the constant RS defined by the reaction $(\varnothing,\varnothing, \{\heartsuit\})$ alone.
        
        By construction, $\res_{\sA} = \res_{\sB}$ when all assignments satisfy $\varphi$. Since the map $\varphi \mapsto (\sA,\sB)$ is computable in polynomial time, deciding if $\res_{\sA} = \res_{\sB}$ is \coNP-hard.
    \end{proof}
\end{teor}
Since in the previous proof the RS $\sB$ is constant, the problem of deciding if the result function of a RS is a non-empty constant is \coNP-complete in unconstrained reaction systems.
In contrast, deciding if a result function is empty can be done in polynomial time.
\begin{cor}
    Given $\sA \in \RS(\infty,\infty)$, it is \coNP-complete to decide if $\res_{\sA}$ is a non-empty constant function; however, deciding if $\res_{\sA} = \varnothing$ is in \Pp.
    \begin{proof}
        The first part of the statement follows directly from the proof of Theorem~\ref{teor: res_A=res_B}. 
        Note that given a RS $\sA = (S,A)$ and a reaction $a = (R_a,I_a,P_a)\in A$, there exists a state $T\subseteq S$ that enables $a$ if and only if $R_a \cap I_a = \varnothing$. 
        Since $\res_{\sA} = \varnothing$ if and only if any reaction is not enabled by all states, we just need to check that $R_a \cap I_a \ne \varnothing$ for all $a \in A$.
    \end{proof}
\end{cor}

We now consider the same problem for inhibitorless RS.
\begin{prop}
    \label{prop: res_A<res_B_inhibitorless}
    Given $\sA=(S,A),\sB = (S,B) \in \RS(\infty, 0)$, if 
    \begin{equation}
        \label{eq: res_A<res_B_inhibitorless}
        \res_{\sA} (R_a) \subseteq \res_{\sB} (R_a) \quad \forall a \in A,
    \end{equation}
    then $\res_{\sA} (T) \subseteq \res_{\sB} (T) $ for all states $T\subseteq S$.
    \begin{proof}
        Let $T\subseteq S$ such that $T\ne R_a$ for all $a\in A$.
        By definition we have
        \begin{equation*}
            \res_{\sA}(T) = \bigcup_{a\in A: R_a\subsetneq T} \res_a(T)
                        = \bigcup_{a\in A: R_a\subsetneq T} \res_a(R_a)
                        = \bigcup_{a\in A: R_a\subsetneq T} \res_{\sA}(R_a).
        \end{equation*}
        By monotonicity of $\res_\sB$, if $ R_a\subseteq T$ then $\res_{\sB} (R_a) \subseteq \res_{\sB} (T)$, so using (\ref{eq: res_A<res_B_inhibitorless}) we obtain
        \begin{equation*}
            \res_{\sA}(T) \subseteq \bigcup_{a\in A: R_a\subsetneq T} \res_{\sB}(R_a) \subseteq \res_{\sB}(T)\,.\qedhere
        \end{equation*}
    \end{proof}
\end{prop}

\begin{cor}
    \label{cor: inibithorless_resA=resB}
     Given $\sA,\sB \in \RS(\infty,0)$ with a common background set $S$, it is in \Pp to decide whether $\res_{\sA} = \res_{\sB}$.
     \begin{proof}
         Applying Proposition~\ref{prop: res_A<res_B_inhibitorless} twice, it is possible to verify in polynomial time that $\res_{\sA}(T) \subseteq \res_{\sB} (T)$ and $\res_{\sB}(T) \subseteq \res_{\sA} (T)$ for all states $T\subseteq S$.
     \end{proof}
\end{cor}
With a proof similar to the one of Proposition~\ref{prop: res_A<res_B_inhibitorless}, we obtain the following result for reactantless RS.
\begin{prop}
    \label{prop: res_A<res_B_reactantless}
    Given $\sA=(S,A), \sB = (S,B) \in \RS(0,\infty)$, if 
    \begin{equation}
        \label{eq: res_A<res_B_reactantless}
        \res_{\sA} (S\setminus I_a) \subseteq \res_{\sB} (S\setminus I_a) \quad \forall a \in A,
    \end{equation}
    then $\res_{\sA} (T) \subseteq \res_{\sB} (T) $ for all states $T\subseteq S$.
    \begin{proof}
        Let $T\subseteq S$ such that $T\ne S\setminus I_a$ for all $a\in A$.
        By definition we have
        \begin{equation*}
            \res_{\sA}(T) = \bigcup_{a\in A: I_a\cap T = \varnothing} \res_a(T)
                        = \bigcup_{a\in A: I_a\cap T = \varnothing} \res_a(S\setminus I_a)
                        = \bigcup_{a\in A: I_a\cap T = \varnothing} \res_{\sA}(S\setminus I_a).
        \end{equation*}
        If $ I_a\cap T = \varnothing$ then $ T \subseteq S\setminus I_a$ and, since $\sB$ is antitone, we have $\res_{\sB} (T) \supseteq \res_{\sB} (S\setminus I_a)$. Using (\ref{eq: res_A<res_B_reactantless}), we obtain
        \begin{equation*}
            \res_{\sA}(T) \subseteq \bigcup_{a\in A: I_a\cap T = \varnothing} \res_{\sB}(S\setminus I_a) \subseteq \res_{\sB}(T)\,.\qedhere
        \end{equation*}
    \end{proof}
\end{prop}

\begin{cor}
    \label{cor: reactantless_resA=resB}
     Given $\sA, \sB \in \RS(0,\infty)$ with a common background set $S$, it is in \Pp to decide whether $\res_{\sA} = \res_{\sB}$.
\end{cor}

\section{Bijective Result Function}
\label{sec: bij_res_A}
In this section, we study the problem of deciding if the result function of a RS is bijective.
This problem is \coNP-complete for $\RS(\infty,\infty)$ \cite[Theorem 7]{formenti2014cycles}.
In this section, we prove that for inhibitorless and reactantless reaction systems, the problem is in \Pp.

\begin{prop}
    \label{prop: |f(T)|=|T|}
    Given $S$ a finite set and $f: 2^S \to 2^S$ monotone and bijective, then $|f(T)| = |T|$ for all $T\subseteq S$.
    \begin{proof}
        Fix $T\subseteq S$ and set $k \coloneqq |T|$, $n\coloneqq |S|$. Given $T_1, T_2\subseteq S$ such that $T_1 \subsetneq T \subsetneq T_2$ then, since $f$ is monotone and injective, it holds
        \begin{equation}
            \label{eq: f(T1)<f(T)<f(T2)}
            f(T_1) \subsetneq f(T) \subsetneq f(T_2).
        \end{equation}
        We can deduce two facts from (\ref{eq: f(T1)<f(T)<f(T2)}) and the injectivity of $f$:
        \begin{enumerate}[(i)]
            \item $f(T)$ strictly contains $2^k-1$ distinct subsets of $S$.
            \item $f(T)$ is strictly contained in $2^{n-k}-1$ distinct subsets of $S$.
        \end{enumerate}
        Now suppose towards a contradiction that $m\coloneqq|f(T)|< k$; then $f(T)$ can strictly contain at most $2^m-1<2^k-1$ different subsets of $S$, contradicting (i).
        On the other hand, if $m>k$ then $f(T)$ is strictly contained in $2^{n-m}-1 > 2^{n-k}-1$ different subsets of $S$, and this contradicts (ii).
        We thus conclude that $m=k$, i.e., $|f(T)| = |T|$.
    \end{proof}
\end{prop}
The first consequence of Proposition~\ref{prop: |f(T)|=|T|} is that bijective monotonic functions are completely determined by their values on the singletons.
\begin{cor}
    \label{cor: f(T)=cup_f(x)}
     Given $S$ a finite set and $f: 2^S \to 2^S$ monotone and bijective,
     then for all $T\subseteq S$
     \begin{equation}
        \label{eq: f(T)=cup_f(x)}
         f(T) = \bigcup_{x\in T} f(\{x\}).
     \end{equation}
     \begin{proof}
         Since $f$ is injective (thus, in particular, it is injective on singletons), we have $|\cup_{x\in T} f(\{x\})| = |T|$; and by Proposition~\ref{prop: |f(T)|=|T|} we have $|\cup_{x\in T} f(\{x\})| = |f(T)|$.
         By monotonicity of $f$, $\cup_{x\in T} f(\{x\}) \subseteq f(T)$; therefore, since the two sets have the same cardinality, they are equal. 
     \end{proof}
\end{cor}

\begin{prop}
    \label{prop: f(T)=cup_f(x)}
    Given $f: 2^S \to 2^S$ injective on singletons such that $f(\varnothing) = \varnothing$, $|f(\{x\})|=1$ for all $x\in S$ and Equation (\ref{eq: f(T)=cup_f(x)}) holds for all $T\subseteq S$, then $f$ is monotone and injective.
    \begin{proof}
        Monotonicity follows directly from Equation (\ref{eq: f(T)=cup_f(x)}). To prove injectivity, consider $T_1\ne T_2$, then there exists $x\in T_1$, $x\notin T_2$. Since $f$ is injective on singletons, $f(\{x\})\ne f(\{y\})$ for all $y\in T_2$. Then $f(\{x\}) \in f(T_1)$ and $f(\{x\})\notin f(T_2)$, so in particular $f(T_1)\ne f(T_2)$.
    \end{proof}
\end{prop}

\begin{rk}
    Given $\sA\in \RS(\infty,0)$ such that $\res_{\sA}$ is injective then $\res_{\sA}$ is additive, by Proposition~\ref{prop: f(T)=cup_f(x)} and Corollary~\ref{cor: f(T)=cup_f(x)}.
    This means that $\sA$ can be $1$-simulated by a reaction system in $\RS(1,0)$ obtained deleting the reactions of $\sA$ with more than two entities in the reactants.
\end{rk}

The following sufficient and necessary conditions for a monotonic function to be bijective follow from Propositions~\ref{prop: |f(T)|=|T|} and~\ref{prop: f(T)=cup_f(x)} and Corollary~\ref{cor: f(T)=cup_f(x)}.
\begin{cor}
    \label{cor: mon_bij_iff}
    Given $\sA= (S,A) \in \RS(\infty,0)$, $\res_{\sA}$ is injective if and only if the following three conditions are satisfied:
    \begin{enumerate}
        \item $\res_{\sA}(\varnothing) = \varnothing$, $|\res_{\sA}(\{x\})| = 1$ for all $x\in S$.
        \item $\res_{\sA}$ is injective on singletons.
        \item for all $(R,\varnothing, P)\in A$, it holds
            $\res_{\sA}(R) = \bigcup_{x\in R} \res_{\sA} (\{x\})$.
    \end{enumerate}
    \begin{proof}
            ($\Rightarrow$) Follows directly from Proposition~\ref{prop: |f(T)|=|T|} and Corollary~\ref{cor: f(T)=cup_f(x)}.
            
            ($\Leftarrow$) Arguing as in the proof of Proposition~\ref{prop: res_A<res_B_inhibitorless}, we obtain that for all $T\subseteq S$,
            \begin{equation*}
                \res_{\sA} (T) = \bigcup_{a\in A: R_a\subseteq T} \res_{\sA}(R_a)
                = \bigcup_{a\in A: R_a\subseteq T} \bigcup_{x\in R_a} \res_{\sA}(\{x\}).
            \end{equation*}
            By Condition 2, every element $x\in T$ belongs to some reaction of the form $(\{x\}, \varnothing, P_x)\in A$ with $|P_x|=1$, so we obtain $\res_{\sA} (T) = \bigcup_{x\in T} \res_{\sA}(\{x\})$.
            The conclusion follows from Proposition~\ref{prop: f(T)=cup_f(x)}.
    \end{proof}
\end{cor}
Given an inhibitorless RS, we can check the three conditions of Corollary~\ref{cor: mon_bij_iff} in polynomial time, obtaining the following.
\begin{cor}
    \label{cor: res_A_bijective_inhibitorless}
    Given $\sA \in \RS(\infty,0)$, deciding whether $\res_{\sA}$ is bijective is in \Pp. 
\end{cor}
Recall that if a function $f:2^S\to 2^S$ is antitone, then $f^2$ is monotone.
Due to this remark, we can decide in polynomial time if the result function of a reactantless RS is bijective.
\begin{cor}
    \label{cor: res_A_bijective_reactantless}
    Given $\sA \in \RS(0,\infty)$, deciding whether $\res_{\sA}$ is bijective is in \Pp.
    \begin{proof}
        Since $\res_{\sA}$ is antitone, $\res_{\sA}^2$ is monotone. Furthermore, $\res_{\sA}$ is injective if and only if $\res_{\sA}^2$ is injective. By Corollary~\ref{cor: res_A_bijective_inhibitorless}, we can check in polynomial time whether $\res_{\sA}^2$ is injective, and since $\res_{\sA}^2(T)$ can be computed in polynomial time from $\res_{\sA}(T)$, the statement follows.
    \end{proof}
\end{cor}

\section{Conclusions}\label{sec:conclusions}
We have determined the computational complexity of an extensive set of decision problems regarding the dynamical behaviour of reactantless and inhibitorless reaction systems. This analysis contributes to providing a more comprehensive understanding of how problem complexity varies across different models. 
Our findings reveal that the simplification of models does not uniformly reduce complexity: most of the analyzed problems retain the same complexity as in the unconstrained model in both reactantless and inhibitorless systems, some become simpler in both the constrained settings, and some others are equally difficult in unconstrained and reactantless systems but become polynomially decidable in inhibitorless systems.

We leave as an open problem to determine the computational complexity of deciding on the existence of a fixed point attractor in an inhibitorless reaction system.
As future directions for extending this work, we also plan to study the complexity of other problems related to the dynamics of resource-bounded reaction systems: for instance, studying cycles and global attractors, similar to what has been done for resource-unbounded systems~\cite{formenti2014cycles}. 
Moreover, it would be interesting to establish the computational complexity of the problems analyzed in this paper in even more constrained classes of reaction systems, such as the special case of inhibitorless reaction systems using only \emph{one} reactant per reaction.
\bibliographystyle{elsarticle-num} 
\bibliography{bibdatabase.bib}




%%\bibitem{}

\end{document}